\documentclass[draftclsnofoot,onecolumn]{IEEEtran}
\usepackage{pifont}
\usepackage{graphicx}
\usepackage{fancyhdr}
\usepackage{amsmath,amssymb,amstext,amsfonts}
\usepackage{slashbox}
\usepackage{bm}
\usepackage{algorithm}
\usepackage{array}
\usepackage{cite}

\usepackage{subfigure}
\usepackage{balance}
\usepackage{algorithmicx}
\usepackage{algpseudocode}

\usepackage{braket}
\usepackage{tabulary}
\usepackage{diagbox}
\usepackage{amsthm}
\usepackage{color}
\makeatletter
\renewcommand{\maketag@@@}[1]{\hbox{\m@th\normalsize\normalfont#1}}%
\makeatother

\newtheorem{theorem}{Theorem}
\newtheorem{lemma}{Lemma}

\begin{document}

\title{Multi-Stage CD-Kennedy Receiver for QPSK Modulated CV-QKD in Turbulent Channels}
\author{Renzhi~Yuan, \IEEEmembership{Member,~IEEE}, Zhixing~Wang, Shouye~Miao, Mufei~Zhao, \IEEEmembership{Member,~IEEE}, Haifeng~Yao,

Bin~Cao, \IEEEmembership{Senior Member,~IEEE}, and Mugen~Peng, \IEEEmembership{Fellow,~IEEE}.
\thanks{
Renzhi Yuan, Zhixing Wang, Shouye~Miao, Bin Cao, and Mugen Peng are with the State Key Laboratory of Networking and Switching Technology, Beijing University of Posts and Telecommunications, Beijing, China; Shouye Miao is also with China United Network Communications Group Company Ltd, Beijing 100033, China; Mufei Zhao is with School of Computer Science and Engineering, Northeastern University, Shenyang, 110819, China; Haifeng~Yao is with School of Optics and Photonics, Beijing Institute of Technology, Beijing 100081, China, and Yangtze Delta Region Academy of Beijing Institute of Technology, Jiaxing 314019, China.}
\thanks{This work is supported by the National Natural Science Foundation of China under No. 62201075.}
\thanks{Corresponding Author: Mugen Peng (pmg@bupt.edu.cn). A journal version of this paper is under peer-review process.}}

\maketitle

\begin{abstract}
Continuous variable-quantum key distribution (CV-QKD) protocols attract increasing attentions in recent years because they enjoy high secret key rate (SKR) and good compatibility with existing optical communication infrastructure. Classical coherent receivers are widely employed in coherent states based CV-QKD protocols, whose detection performance is bounded by the standard quantum limit (SQL). Recently, quantum receivers based on displacement operators are experimentally demonstrated with detection performance outperforming the SQL in various practical conditions. However, potential applications of quantum receivers in CV-QKD protocols under turbulent channels are still not well explored, while practical CV-QKD protocols must survive from the atmospheric turbulence in satellite-to-ground optical communication links. In this paper, we consider the possibility of using a quantum receiver called multi-stage CD-Kennedy receiver to enhance the SKR performance of a quadrature phase shift keying (QPSK) modulated CV-QKD protocol in turbulent channels. We first derive the error probability of the multi-stage CD-Kennedy receiver for detecting QPSK signals in turbulent channels and further propose three types of multi-stage CD-Kennedy receiver with different displacement choices, i.e., the Type-I, Type-II, and Type-III receivers. Then we derive the SKR of a QPSK modulated CV-QKD protocol using the multi-stage CD-Kennedy receiver and post-selection strategy in turbulent channels. Numerical results show that the multi-stage CD-Kennedy receiver can outperform the classical coherent receiver in turbulent channels in terms of both error probability and SKR performance and the Type-II receiver can tolerate worse channel conditions compared with Type-I and Type-III receivers in terms of error probability performance. Besides, we also demonstrate that the Type-III receiver can be a good trade-off choice between error probability and SKR performance in all range of received signal strengths and it can provide significant improvements in both error probability and SKR gains over the classical homodyne receiver in weak links. Our work shed a light on the practical implementation of quantum receivers in satellite-to-ground optical communication links with turbulent channels.
\end{abstract}

\begin{IEEEkeywords}
CV-QKD, post-selection strategy, quantum receiver, QPSK, turbulent channels
\end{IEEEkeywords}

%
\IEEEpeerreviewmaketitle

\section{Introduction}
\subsection{Background and Motivation}
Quantum key distribution (QKD) is widely recognized as a promising solution to endogenous security challenges within the envisioned space-air-ground integrated network, offering unconditional security guarantees at the physical layer by leveraging the fundamental principles of quantum mechanics \cite{wang2021analysis}. Notably, continuous variable QKD (CV-QKD) \cite{grosshans2002continuous,silberhorn2002continuous,grosshans2003quantum} exhibits superior secret key rates and enhanced compatibility with existing optical communication infrastructures compared to its discrete variable counterpart (DV-QKD) \cite{pirandola2020advances}. These advantages have consequently garnered significant research interest in recent years \cite{ghalaii2023continuous,liu2025otfs,yao2025continuous}.

Conventional CV-QKD protocols predominantly utilize classical coherent receivers for bit recovery \cite{grosshans2002continuous,silberhorn2002continuous,grosshans2003quantum}, though their performance remains fundamentally constrained by the standard quantum limit (SQL) \cite{wittmann2010demonstration}. To overcome this limitation, the generalized Kennedy receiver, a quantum-enhanced detection scheme, has been incorporated into post-selection-based CV-QKD systems, demonstrating capability to surpass the SQL and elevate secret key rates (SKRs) \cite{wittmann2010demonstration}. Empirical studies have further validated that such quantum receivers achieve higher SKRs than classical counterparts under both individual and collective attack scenarios \cite{wittmann2010demonstration,zhao2020security,zhao2021post}. However, existing implementations of quantum receiver-enhanced CV-QKD are exclusively confined to loss-dominated channels. Notably, within space-air-ground integrated networks, satellite-to-ground links constitute the most critical and vulnerable segment due to atmospheric turbulence effects \cite{zhu2002free}. Some preliminary efforts have been made to implement quantum receivers in turbulent channels for CV-QKD protocols employing binary phase shift keying (BPSK) modulation \cite{miao2025generalized}. However, quadrature phase shift keying (QPSK) represents another widely used modulation scheme in satellite laser communications and it enjoys higher modulation efficiency compared with BPSK modulation. To the best of our knowledge, the study of quantum receivers for discriminating QPSK signals in turbulent channels remains unexplored. Besides, the application research of quantum-enhanced receivers to QPSK modulated CV-QKD protocols over turbulent channels is also urgently needed.

\subsection{Related Works}

The CV-QKD protocol based on coherent states was initially proposed by Grosshans and Grangier \cite{grosshans2002continuous}, which guarantees security via the no-cloning theorem yet suffers from a 3dB loss limit. Post-selection strategies subsequently overcame this limitation by selectively retaining high-information transmissions, enabling operation in high-loss environments without quantum resources like entanglement \cite{silberhorn2002continuous}. Conventional implementations of CV-QKD mainly employ classical coherent receivers, whose detection performance is fundamentally bounded by the SQL \cite{grosshans2002continuous,silberhorn2002continuous,grosshans2003quantum}. Quantum receivers utilizing optimal state discrimination principles \cite{burenkov2021practical} pioneered by Helstrom \cite{helstrom1969quantum,helstrom1970quantum} surpass this barrier.

Early studies on quantum receivers mainly focused on the discrimination of binary modulated signals \cite{burenkov2021practical}. For example, the optimal decision theory for discriminating two quantum states was established by Helstrom in the late 1960s, known as the Helstrom bound. The first realizable quantum receiver for discriminating BPSK modulated coherent states was proposed by Kennedy in 1972 \cite{kennedy1973near}, where a displacement operation was employed to cancel one of the transmitted coherent state and an on/off photodetector was employed to decide the transmitted bit. In 1973, Dolinar generalized Kennedy's receiver structure by introducing realtime displacement control with feedback links, which can well approach the Helstrom bound. However, the experimental realization of Dolinar's receiver has not been achieved until 2007 by Cook et al \cite{cook2007optical}. Subsequent advances include displacement optimization \cite{wittmann2008demonstration, takeoka2008discrimination}, photon-number-resolving detection \cite{wittmann2010demonstration}, and noise resilience \cite{yuan2020kennedy,suo2025phase,yuan2020optimally}.  Crucially, when integrated with CV-QKD protocols \cite{wittmann2010demonstration,zhao2020security,zhao2021post}, these receivers demonstrate enhanced secret key rates (SKRs) in lossy channels.

The extension of generalized Kennedy receiver to multiple phase shift keying (MPSK) modulated coherent states attracted great interests because of the high modulation efficiency of MPSK signals \cite{burenkov2021practical,zhao2023optimally}. Though the optimal decision theory for MPSK signals was given by Helstrom's theory \cite{helstrom1969quantum}, the first realizable quantum receiver for discriminating QPSK signals was proposed by Bondurant in 1993 \cite{bondurant1993near}, where a multi-stage receiving structure was used to cancel one of the transmitted signals. This multi-stage receiving structure was extended to discriminating MPSK modulated coherent states by Becerra et al in 2011 and was experimentally demonstrated for QPSK signals \cite{becerra2011m}. Later in 2012, Shuro et al studied the multi-stage quantum receiver by using adaptive measurement with feedforward links and on/off photodetectors \cite{izumi2012displacement}. This receiver was improved by replacing on/off photodetectors with photon number resolving detector (PNRD) and was experimentally demonstrated in 2013 \cite{izumi2013quantum,li2013suppressing,becerra2013experimental}. The SQL surpassing performance of the multi-stage quantum receiver with adaptive measurement and PNRD was further extended to power levels compatible with state-of-the-art optical communication systems in 2014 \cite{becerra2014photon}, demonstrating the potential applications of quantum receivers to realistic coherent optical communication systems. In 2020, an experimental demonstration of the multi-stage quantum receiver at telecommunication wavelength was achieved by using a high performance single photon detector \cite{izumi2020experimental}. Some attempts of combining deep learning‌ methods with quantum receivers for discriminating multiple coherent states were also reported in recent years \cite{qu2022learnable,cui2022quantum}. Besides, the multi-stage quantum receiver with adaptive measurement was also employed to improve the SKR of QPSK modulated CV-QKD \cite{liao2018long,zhao2024security}.

However, existing quantum receiver studies almost focused on turbulence-free scenarios. The extension of quantum receiver to turbulent channel is of great importance because atmospheric turbulence dominates satellite-to-ground link vulnerability in space-air-ground integrated networks. To mitigate the turbulent influence on discriminating BPSK modulated coherent states, a conditionally-dynamic based Kennedy (CD-Kennedy) receiver was proposed in a preliminary study \cite{yuan2020free} and was recently combined with the BPSK modulated CV-QKD protocol in turbulent channels \cite{miao2025generalized}. However, QPSK represents another widely used modulation scheme in satellite laser communications and the quantum receiving structure for QPSK signals is much complicated compared with BPSK counterpart. Therefore, the study of quantum receivers for discriminating QPSK signals in turbulent channel is an essential yet unexplored research frontier. Besides, critical research is also required on the application of quantum receivers to QPSK-modulated CV-QKD protocols deployed in turbulent channels.

\subsection{Contributions}
In this paper, we propose a multi-stage CD-Kennedy receiver for discriminating QPSK signals in turbulent channels and consider the possibility of using multi-stage CD-Kennedy receiver to enhance the SKR performance of a QPSK modulated CV-QKD protocol in turbulent channels. We first derive the error probability of the multi-stage CD-Kennedy receiver for detecting QPSK signals in turbulent channels and propose three types of multi-stage CD-Kennedy receiver with different displacement choices. Then we derive the SKR of a QPSK modulated CV-QKD protocol by using the multi-stage CD-Kennedy receiver in turbulent channels, where a post-selection strategy on both the receiving number of photons at all stages and the channel transmittance is used. Numerical simulations are conducted to verify the both the error probability and SKR performance of the proposed multi-stage CD-Kennedy receiver in turbulent channels. The major contribution of this work can be summarized as follows:
\begin{itemize}
\item For the first time we proposed a multi-stage CD-Kennedy receiver for discriminating QPSK signals in turbulent channels and derived the error probability of the multi-stage CD-Kennedy receiver. Besides, we further proposed three types of multi-stage CD-Kennedy receiver, i.e., the Type-I, Type-II, and Type-III receivers, with different displacement choices to improve the receiving performance.
\item For the first time we employed the multi-stage CD-Kennedy receiver to improve the SKR performance of the QPSK modulated CV-QKD protocol in turbulent channels and derived the SKR of the multi-stage CD-Kennedy receiver  using a post-selection strategy on both the receiving number of photons at all stages and the channel transmittance.
\item We demonstrated that the proposed three types of multi-stage CD-Kennedy receiver can outperform the classical coherent receiver in turbulent channels in terms of both error probability and SKR performance. Besides, the Type-II receiver can tolerate worse channel conditions compared with Type-I and Type-III receivers in terms of error probability performance. However, when the SKR performance is considered, the advantage of Type-II receiver vanishes under small received signal strengths.
\item We also demonstrated that the Type-III receiver is a good trade-off choice between error probability and SKR performance in all range of received signal strengths. Besides, the Type-III receiver can provide significant improvements in both error probability and SKR gains over the classical homodyne receiver in weak links. Our work shed a light on the practical implementation of quantum receivers in satellite-to-ground optical communication links.
\end{itemize}

The rest of this paper is organized as follows: we first present the structure of multi-stage CD-Kennedy receiver and derive the error probability of the multi-stage CD-Kennedy receiver under turbulent channels in Section \ref{multi-stage_CD-Kennedy_BER}; then we study the SKR of the QPSK modulated CV-QKD protocol by using multi-stage CD-Kennedy receiver with post-selection strategy under turbulent channels in Section \ref{CV-QKD_with_CD-Kennedy}; some numerical results are presented to explore the error probability and SKR performance of multi-stage CD-Kennedy receiver in turbulent channels in Section \ref{Numerical_Results}; and finally we conclude our work in Section \ref{Conclusion}.

\begin{figure*}
\centering
\includegraphics[width=0.95\linewidth]{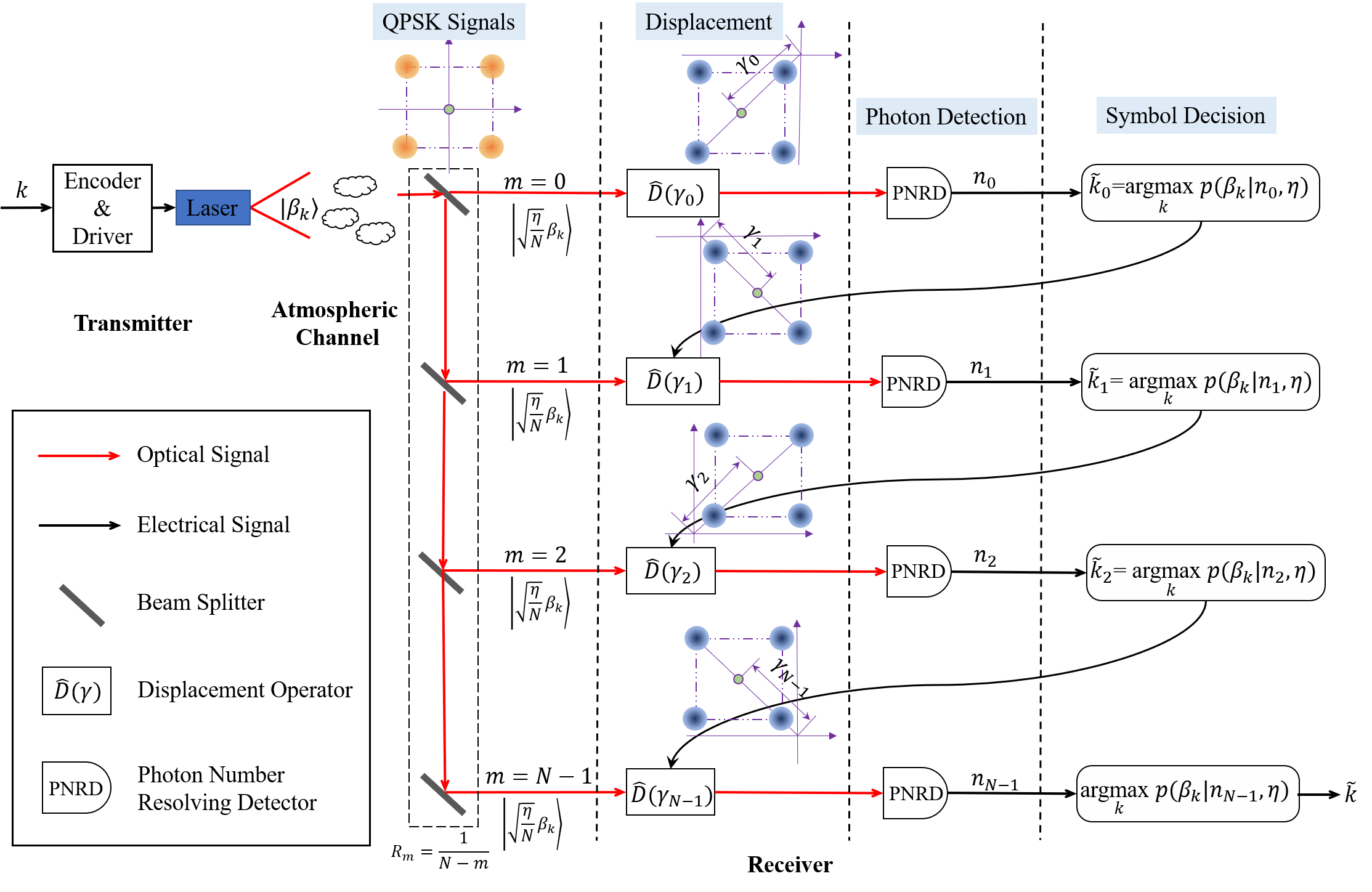}
\caption{Multi-stage CD-Kennedy receiver with feedforward for discriminating QPSK signals}
\label{system_set}
\end{figure*}

\section{Multi-Stage CD-Kennedy Receiver in Turbulent Channel}\label{multi-stage_CD-Kennedy_BER}

We consider a QPSK modulated optical communication system employing a multi-stage CD-Kennedy receiver with adaptive feedforward in this paper, as shown in Fig. \ref{system_set}. The transmitted symbol $k \in \{0,1,2,3\}$ is encoded by a coherent state $\ket{\beta_k}=\ket{\beta e^{i\frac{\pi}{4}(2k+1)}}$ with $k=0,1,2,3$. Without loss of generality, we adopt $\beta$ as real number. After passing through a turbulent channel, the transmitted signals are received by a multi-stage CD-Kennedy receiver with adaptive feedforwards.

In a CD-Kennedy receiver, the transmittance $\eta$ of the turbulent channel is first estimated by some pilot symbols and then used to decide the displacement values of the receiver \cite{yuan2020free}. Specifically, the coherent state arriving at the receiver $\ket{\sqrt{\eta}\beta_k}$ is divided into $N$ branches, where the reflection of the beam splitter (BS) at the $m$th branch is set as $R_m=\frac{1}{N-m}$ such that the optical intensity is equally divided into $N$ parts and the input state at each branch becomes $\ket{\psi_k}\triangleq \ket{\sqrt{\frac{\eta}{N}}\beta_k}$ with $k=0,1,2,3$. A displacement operation $\hat{D}(\gamma_m)$ is employed to displace the input state at the $m$th branch and a PNRD is used to detect the number of photons $n_m$ in the displaced state. At the first branch with $m=0$, the displacement operator $\hat{D}(\gamma_0)$ is chosen to displace the received state along the inverse direction of the transmitted state with the largest prior probability. Then a maximum a posteriori (MAP) decision rule on the detected photon numbers $n_m$ is employed to decide the most likely transmitted state inferred from the $m$th branch. The calculated posteriori probabilities at the $m$th branch is used as the prior probability of the $(m+1)$th branch and the displacement operator $\hat{D}(\gamma_{m+1})$ is chosen to displace the received state along the inverse direction of the quantum state decided at the $m$th branch. After $N$ stages, the final output at the last branch is adopted as the estimate of the transmitted symbol $\tilde{k}$. In the following, we derive the error probability of this communication system.

\subsection{Coherent States Transmission in Turbulent Channel}
The quantum state of a laser signal can be expressed in a coherent state $\ket{\alpha}$ with $\alpha\in \mathbb{C}$ and $\mathbb{C}$ is the field of complex number. By using the Fock basis of the Hilbert space, which consists of all number states (or Fock states) ${\{\ket{n},n=0, 1, 2, ...\}}$, we can expand the coherent sate
$\ket{\alpha}$ as \cite{glauber1963coherent}
\begin{equation}\label{coherent_state}
{\ket{\alpha}} = \sum_{n=0}^{\infty}e^{-\frac{1}{2}{|\alpha|}^2}\frac{{\alpha}^n}{\sqrt{n!}}\ket{n},
\end{equation}
\noindent where $|\alpha|^2$ is the average number of photons contained in the coherent state $\ket{\alpha}$. According to Glauber's theory, all the coherent states ${\{\ket{\alpha}, \alpha \in \mathbb{C}\}}$ form an overcomplete basis of Hilbert space; and therefore, any quantum state with a density operator ${\hat{\rho}}$ in this Hilbert space could be decomposed by coherent states as \cite{glauber1963coherent}
\begin{equation}
\hat{\rho} = \int_{\alpha}P(\alpha)\ket{\alpha}\bra{\alpha}\mathrm{d}^2\alpha,
\end{equation}
\noindent where $\mathrm{d}^2\alpha=\mathrm{d}Re(\alpha)\mathrm{d}Im(\alpha)$ and ${P(\alpha)}$ is the $P$-function of the density operator $\hat{\rho}$. This representation of density operator is called the $P$-representation \cite{glauber1963coherent}. Specifically, the $P$-function of a coherent state $\ket{\beta_l}$ can be expressed as a Dirac delta function $\delta^2(\alpha-\beta_l)\triangleq \delta\left(Re(\alpha)-Re(\beta_l)\right)\delta\left(Im(\alpha)-Im(\beta_l)\right)$.

The relation between the input and output $P$-functions of turbulent channel was first derived by Semenov \cite{semenov2009quantum,yuan2020closed} as
\begin{equation}\label{Equa:TurbulenceChannel}
P_{out}(\alpha)=\int_{\eta}\frac{p(\eta)}{\eta}P_{in}(\frac{\alpha}{\sqrt{\eta}})\mathrm{d}\eta,
\end{equation}
\noindent where $\eta$ is the transmittance of the channel; $p(\eta)$ is the probability density function (PDF) of the transmittance; $P_{in}(\alpha)$ is the input $P$-function and $P_{out}(\alpha)$ is the output $P$-function of the turbulent channel. We adopt a log-normal distributed turbulent channel in this paper, where the PDF of the transmittance $\eta$ satisfies the following log-normal PDF:
\begin{equation}
\label{Equa:LNdistribution}
\begin{aligned}
p(\eta)&=\frac{1}{\eta\sqrt{2\pi\sigma_t^2}}e^{-\frac{(\ln (\eta/\eta_0)+0.5\sigma_t^2)^2}{2\sigma_t^2}},
\end{aligned}
\end{equation}
\noindent where $\sigma_t^2$ is the variance of $\ln \eta$ and $\eta_0\triangleq \text{E}[\eta] \in [0,1]$ is the expectation of $\eta$.

Substituting the $P$-function of coherent state into \eqref{Equa:TurbulenceChannel}, we can obtain the output $P$-function of the turbulent channel as \cite{yuan2020free}
\begin{equation}\label{Equa:PforTurbulence_ith}
\begin{aligned}
P_{tur}(\alpha)&=\int_{\eta}\frac{p(\eta)}{\eta}\delta^2\left(\frac{\alpha}{\sqrt{\eta}}-\beta_k\right)\mathrm{d}\eta.
\end{aligned}
\end{equation}

\subsection{Error Probability of Multi-Stage CD-Kennedy Receiver for QPSK Signals in Turbulent Channel}

We adopt the CD-Kennedy receiver structure proposed in \cite{yuan2020free} to mitigate the influence of turbulence, where the displacement value $\gamma_m$ of the $m$th branch is dynamically conditioned on both the transmittance $\eta$ and the estimated symbol of the $(m-1)$th branch. The transmittance $\eta$ is estimated by pilot bits under slow-varying turbulent channels \cite{yuan2020free,zhu2002free}. In this context, the transmittance $\eta$ can be regarded as a fixing value for a given time period. Then the $P$-function of the input state at the $m$th branch given transmittance $\eta$ and transmitted signal $\ket{\beta_k}$ can be obtained as
\begin{equation}\label{Equa:PforTurbulence_ith_given_eta_i}
\begin{aligned}
P_{tur}(\alpha|\eta,\beta_k)&=\delta^2\left(\alpha-\psi_k\right),
\end{aligned}
\end{equation}
\noindent which corresponds to a coherent state $\ket{\psi_k}$.

Then we use a displacement operator $\hat{D}(\gamma_m)$ to displace the input state $\ket{\psi_k}$. In practical implementation, the displacement operator $\hat{D}(\gamma_m)$ can be achieved by combining the input coherent state with an local oscillator (LO) with $\ket{\alpha_{LO}}=\ket{\sqrt{\frac{\tau}{1-\tau}}\gamma_m}$ using a high transmittance BS with transmittance $\tau\to 1$. Then the displaced state can be approximated as $\ket{\psi_k+\gamma_m}$.

Then a PNRD is employed to detect the number of photons in the displaced state $\ket{\psi_k+\gamma_m}$. An ideal PNRD by a set of positive-operator
valued measure (POVM) operators $\{\hat{M}_n\}_{n=0}^\infty$, where the operator $\hat{M}_n=\ket{n}\bra{n}$. Then the detection probability of $n_m$ photons at the $m$th branch given transmitted symbol $\ket{\beta_k}$, turbulent transmittance $\eta$, and displacement value $\gamma_m$ can be obtained as
\begin{equation}\label{Equa:photon_distribution_given_eta_i}
\begin{aligned}
p(n_m|\beta_k,\eta,\gamma_m)&=\left\langle \psi_k+\gamma_m\right| \hat{M}_n \left|\psi_k+\gamma_m\right\rangle\\
&=\frac{e^{-N_{k,m}}}{n_m!}N_{k,m}^{n_m},
\end{aligned}
\end{equation}
\noindent where $N_{k,m}$ is the average number of photons contained in displaced state $\ket{\psi_k+\gamma_m}$, i.e,
\begin{equation}\label{lambda_m}
\begin{aligned}
N_{k,m}&=\left(\frac{\eta N_s}{N}+|\gamma_m|^2+2\sqrt{\frac{\eta N_s}{N}}|\gamma_m|\right.\\
&\quad\quad\quad\quad\quad \times \cos\big(\arg(\beta_k)-\arg(\gamma_m) \big)\bigg),
\end{aligned}
\end{equation}
\noindent where $N_s\triangleq \beta^2$ is the average signal photons.

An MAP decision rule is used to decide the transmitted symbol at the $m$th branch as
\begin{equation}\label{k_m}
\begin{aligned}
\tilde{k}_m&=\mathop{\arg\max}_{k} \ p(\beta_k|n_{m},\eta,\gamma_m),
\end{aligned}
\end{equation}
\noindent where $p(\beta_k|n_{m},\eta,\gamma_m)$ is the posterior probability of transmitting $\beta_k$ when $n_m$ photons are detected; and $p(\beta_k|n_{m},\eta,\gamma_m)$ is given by
\begin{equation}\label{posterior_prob}
\begin{aligned}
p(\beta_k|n_{m},\eta,\gamma_m)=\frac{p_m(\beta_k) p(n_{m}|\beta_k,\eta,\gamma_m)}{\sum_{k=0}^3 p_m(\beta_k) p(n_{m}|\beta_k,\eta,\gamma_m)},
\end{aligned}
\end{equation}
\noindent where $p_m(\beta_k)$ is the prior probability of the $m$th branch determined by the posterior probability of the $(m-1)$th branch as
\begin{equation}\label{p_m_beta_k}
p_m(\beta_k)=
\begin{cases}
p(\beta_k|n_{m-1},\eta,\gamma_{m-1}), \quad m=1,2,\cdots,N-1\\
p_k,\quad m=0,
\end{cases}
\end{equation}
\noindent and where $p_k$ is the prior probability of transmitting symbol $k$ at the transmitter. Without loss of generality, we adopt equal prior probabilities with $p_k=\frac{1}{4}$ for $k=0,1,2,3$.

Then the posterior probabilities $p(\beta_k|n_m,\eta)$ at the $m$th branch can be further obtained by recursively using \eqref{posterior_prob} and we have
\begin{equation}\label{posterior_prob_last}
\begin{aligned}
p(\beta_k|n_{m},\eta,\gamma_m)=\frac{p_{k}\prod_{i=0}^{m} p(n_{i}|\beta_k,\eta,\gamma_i)}{\sum_{k=0}^3 p_{k} \prod_{i=0}^{m} p(n_{i}|\beta_k,\eta,\gamma_i)}.
\end{aligned}
\end{equation}

Therefore, the output symbol of the $m$th branch can be obtained as
\begin{equation}\label{k_estimate}
\begin{aligned}
\tilde{k}_m&=\mathop{\arg\max}_{k} \ \frac{p_{k}\prod_{i=0}^{m} p(n_{i}|\beta_k,\eta,\gamma_i)}{\sum_{k=0}^3 p_{k} \prod_{i=0}^{m} p(n_{i}|\beta_k,\eta,\gamma_i)}.
\end{aligned}
\end{equation}

The final output symbol is $\tilde{k}=\tilde{k}_{N-1}$. Then the error probability given transmittance $\eta$ and displacement values $\bm{\gamma}\triangleq \{\gamma_0,\gamma_1,\cdots,\gamma_{N-1}\}$ can be obtained as  (see Appendix \ref{Appendix P_e})
\begin{equation}\label{P_e_eta_mid}
\begin{aligned}
P_e(\eta,\bm{\gamma})&=1-\sum_{k=0}^3\sum_{\bm{n}\in Q_k} p_k p(\bm{n}|\beta_k,\eta,\bm{\gamma}),
\end{aligned}
\end{equation}
\noindent where $\bm{n}\triangleq \{n_0,n_1,\cdots,n_{N-1}\}$ and $Q_k\triangleq \{\bm{n}|\tilde{k}=k\}$ is the decision areas on all possible combinations of output photon numbers $\bm{n}$ for symbol $k$; $p(\bm{n}|\beta_k,\eta,\bm{\gamma})$ is the probability of obtaining $\bm{n}$ output photons when $\beta_k$ is transmitted under transmittance $\eta$ and displacement values $\bm{\gamma}$. According to the independent detection properties of all the branches, we can express $p(\bm{n}|\beta_k,\eta,\bm{\gamma})$ as
\begin{equation}\label{p_n_beta_k_eta}
\begin{aligned}
p(\bm{n}|\beta_k,\eta,\bm{\gamma})=\prod_{m=0}^{N-1} p(n_{m}|\beta_k,\eta,\gamma_m),
\end{aligned}
\end{equation}
\noindent where $p(n_{m}|\beta_k,\eta,\gamma_m)$ is given in \eqref{Equa:photon_distribution_given_eta_i}. By substituting \eqref{p_n_beta_k_eta} into \eqref{P_e_eta_mid}, we can obtain the error probability as
\begin{equation}\label{P_e_eta}
\begin{aligned}
P_e(\eta,\bm{\gamma})&=1-\sum_{k=0}^3\sum_{\bm{n}\in Q_k} p_k \prod_{m=0}^{N-1} p(n_{m}|\beta_k,\eta,\gamma_m).
\end{aligned}
\end{equation}

Finally, the error probability over the turbulent channel can be expressed as
\begin{equation}\label{P_e_CD_Kenndy}
P_e(\bm{\gamma})=1-\int_\eta p(\eta)\sum_{k=0}^3\sum_{\bm{n}\in Q_k}p_k\prod_{m=0}^{N-1} p(n_{m}|\beta_k,\eta,\gamma_m) \mathrm{d} \eta,
\end{equation}
\noindent where $p(\eta)$ is the PDF of turbulent channel transmittance given in \eqref{Equa:LNdistribution}.

\subsection{Three Types of Multi-Stage CD-Kennedy Receiver}
The error probability is conditioned on the displacement values $\bm{\gamma}$. Because the displacement operator $\hat{D}(\gamma_m)$ is always along with one of the transmitted signals, for simplicity we can define a displacement ratio as
\begin{equation}\label{alpha_r}
\alpha_m\triangleq \frac{|\gamma_m|}{|\psi_k|}.
\end{equation}

Then the error probability is decided by the displacement ratios $\alpha_0,\alpha_1,\cdots,\alpha_{N-1}$. According to the chosen strategy of the displacement ratios $\alpha_0,\alpha_1,\cdots,\alpha_{N-1}$, here we consider three types of multi-stage CD-Kennedy receivers.

\subsubsection{\textbf{Type-I Receiver (No Optimization)}}
In Type I receiver, the displacement value $\gamma_m$ is chosen to cancel the state with maximum prior probability at each stage, i.e., we have $\alpha_0=\alpha_1=\cdots=\alpha_{N-1}\triangleq \alpha_r=1$. Then we can express $\gamma_m$ as
\begin{equation}\label{gamma_m}
\gamma_m=-\sqrt{\frac{\eta}{N}}\beta_{\tilde{k}_{m-1}},
\end{equation}
\noindent where $\tilde{k}_m$ is given in \eqref{k_m}; and for $m=0$, we can randomly choose $\gamma_0$ to cancel one of the transmitted symbol. Without loss of generality, we set $\gamma_0=-\sqrt{\frac{\eta}{N}}\beta_{0}$. In this context, we can further simplify $N_{k,m}$ as
\begin{equation}\label{lambda_m_1}
N_{k,m}=\frac{2\eta N_s}{N}\left(1+\cos \left(\frac{\pi}{2}(k-\tilde{k}_{m-1})\right)\right).
\end{equation}

\subsubsection{\textbf{Type-II Receiver (Global Optimization)}}
In Type-II receiver, we choose $\alpha_0=\alpha_1=\cdots=\alpha_{N-1}\triangleq \alpha_g$ and optimize the global displacement ratio $\alpha_g$ as
\begin{equation}\label{alpha_g_opt}
\begin{aligned}
\alpha_g^* = \mathop{\arg\min}_{\alpha_g}& \  P_e(\bm{\gamma})\\
\text{s. t.}& \ \gamma_m=-\alpha_g \sqrt{\frac{\eta}{N}}\beta_{\tilde{k}_{m-1}}.
\end{aligned}
\end{equation}
\noindent where $\alpha_g^*$ is the globally optimized displacement radio. Here we use brute-force search to solve the optimization problem in \eqref{alpha_g_opt}.

\subsubsection{\textbf{Type-III Receiver (Optimization on $\alpha_0$)}}

\begin{figure}
\centering
\includegraphics[width=0.45\textwidth]{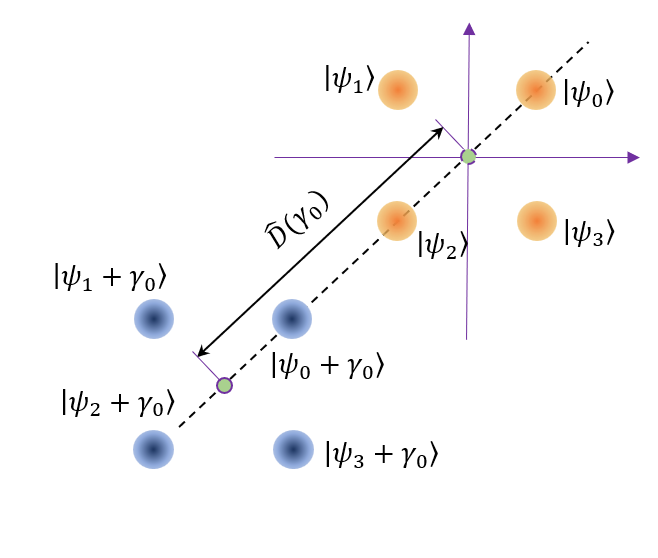}
\caption{Displacement operation on the input QPSK signals at the first stage}
\label{displacement}
\end{figure}

In Type-III receiver, we choose $\alpha_1=\alpha_2=\cdots=\alpha_{N-1}\triangleq 1$ and optimize the displacement ratio $\alpha_0$ only. This is because the decision is made stage by stage and thus the error probability in the first stage can greatly affect the final error probability. Without loss of generality, we choose to displace the input state along with $\ket{\beta_0}$, as shown in Fig. \ref{displacement}, where the displacement value is given by
\begin{equation}
\begin{aligned}
\gamma_0&=-\alpha_0 \sqrt{\frac{\eta}{N}}\beta_{0}\\
&=-\alpha_0 \sqrt{\frac{\eta}{N}}\beta e^{i\frac{\pi}{4}}.
\end{aligned}
\end{equation}

Then the optimized displacement ratio $\alpha_0^*$ can be obtained by minimizing the error probability of the MAP detection at the first stage as
\begin{equation}\label{alpha_0_optimization}
\begin{aligned}
\alpha_0^* = \mathop{\arg\min}_{\alpha_0}& \  P_{e,0}(\alpha_0),
\end{aligned}
\end{equation}
\noindent where the error probability of at the first stage is given by
\begin{equation}\label{P_e_0}
\begin{aligned}
P_{e,0}(\alpha_0)=1-\sum_{k=0}^3P_r(\tilde{k}_0=k|k),
\end{aligned}
\end{equation}
\noindent and where $\tilde{k}_0$ is obtained by the MAP detection given in \eqref{k_m}.

It is challenging to solve the optimization \eqref{alpha_0_optimization} directly because the error probability in \eqref{P_e_0} is not in an analytical expression. Fortunately, the following theorem guarantees that the MAP detection on $\tilde{k}_0$ is equivalent to a multiple-threshold rank detection.

\begin{theorem}
The MAP detection on the QPSK modulated signals at the first stage equals a multiple-threshold rank detection with decision rule given by (see Appendix \ref{multi-threshold_rank_detection}):
\begin{equation}\label{rank_detection}
\begin{aligned}
\tilde{k}_0=
\begin{cases}
0, \ \text{if } n_0\leq n_{th,0},\\
1 \text{ or } 3,   \ \text{if }  n_{th,0}<n_0\leq n_{th,1},\\
2, \ \text{if }   n_0> n_{th,1},
\end{cases}
\end{aligned}
\end{equation}
\noindent where $n_{th,0}$ and $ n_{th,1}$ are two thresholds given by
 \begin{equation}\label{thresholds}
\begin{cases}
n_{th,0}=\big\lfloor \frac{2\alpha_0N_0}{\ln (1+\alpha_0^2)-\ln(1-\alpha_0)^2} \big\rfloor,\\
n_{th,1}=\big\lfloor \frac{2\alpha_0N_0}{\ln(1+\alpha_0)^2-\ln (1+\alpha_0^2)} \big\rfloor,
\end{cases}
\end{equation}
\noindent and where $N_{0}\triangleq |\psi_k|^2=\frac{\eta N_s}{N}$ is the input signal strength at the first stage.
\end{theorem}

Then the error probability in \eqref{P_e_0} can be rewritten as the following analytical form:
\begin{equation}\label{P_e_0_rank_detection}
\begin{aligned}
P_{e,0}(\alpha_0)&=\frac{3}{4}-\frac{1}{4}\left(\sum_{n=0}^{n_{th,0}}\frac{e^{-N_{0,0}}}{n!}N_{0,0}^n
+\!\!\sum_{n=n_{th,0}+1}^{n_{th,1}}\!\!\frac{e^{-N_{1,0}}}{n!}N_{1,0}^n\right.\\
&\quad \quad\quad\quad\quad\quad\quad\quad \left.-\sum_{n=0}^{n_{th,1}}\frac{e^{-N_{2,0}}}{n!}N_{2,0}^n\right),
\end{aligned}
\end{equation}
\noindent where
\begin{equation}
\begin{cases}
N_{0,0}=(1-\alpha_0)^2N_0, \\
N_{1,0}=(1+\alpha_0^2)N_0, \\
N_{2,0}=(1+\alpha_0)^2N_0.
\end{cases}
\end{equation}

Different from the Type-II receiver, the displacement ratio $\alpha_0$ in Type-III receiver only depends on the input signal strength $N_{0}$ at the first stage. Therefore, we can obtain the optimized displacement ratio $\alpha_0^*$ for minimizing $P_{e,0}$ under different input signal strength $N_{0}$ in advance and save $\alpha_0^*$ as a lookup table, which can greatly simplify the computational complexity.

\subsection{Error Probability of Homodyne Receiver and Helstrom Bound for QPSK Signals in Turbulent Channel}
We use a classical homodyne receiver with SQL and also the Helstrom bound for error probability performance comparison.

\subsubsection{Homodyne Receiver}
The output of the classical coherent receiver given transmittance $\eta$ and transmitted symbol $\ket{\beta_k}$ can be modeled as a Gaussian distributed random variable. Then the error probability for detecting QPSK signals given transmittance $\eta$ can be obtained as
\begin{equation}
p_e^{HD}(\eta)=1-\frac{1}{\pi}\int_0^\infty\int_{-\frac{\pi}{4}}^{\frac{\pi}{4}}e^{-|re^{i\theta}-\sqrt{\eta N_s}|^2}r\mathrm{d}r\mathrm{d}\theta.
\end{equation}

Similarly, the error probability of the classical coherent receiver over turbulent channel can be obtained as
\begin{equation}
P_e^{HD}=1-\frac{1}{\pi} \int_0^\infty\int_{-\frac{\pi}{4}}^{\frac{\pi}{4}}\int_\eta p(\eta)e^{-|re^{i\theta}-\sqrt{\eta N_s}|^2}r\mathrm{d}r\mathrm{d}\theta\mathrm{d}\eta.
\end{equation}

\subsubsection{Helstrom Bound}
The Helstrom bound for discriminating QPSK signals given transmittance $\eta$ is given by
\begin{equation}
P_e^{Hel}(\eta)=1-\frac{1}{16}\left(\sum_{k=0}^3\sqrt{\lambda_{H,k}}\right)^2,
\end{equation}
\noindent where $\lambda_{H,0}$, $\lambda_{H,1}$, $\lambda_{H,2}$, and $\lambda_{H,3}$ are given by
\begin{equation}
\begin{cases}
\lambda_{H,0}=2e^{-\eta N_s}(\cosh (\eta N_s)+\cos (\eta N_s)),\\
\lambda_{H,1}=2e^{-\eta N_s}(\sinh (\eta N_s)-\sin (\eta N_s)),\\
\lambda_{H,2}=2e^{-\eta N_s}(\cosh (\eta N_s)-\cos (\eta N_s)),\\
\lambda_{H,3}=2e^{-\eta N_s}(\sinh (\eta N_s)+\sin (\eta N_s)).\\
\end{cases}
\end{equation}

Then the Helstrom bound for discriminating QPSK signals over turbulent channel can be obtained as
\begin{equation}
P_e^{Hel}=1-\frac{1}{16}\int_0^\infty\left(\sum_{k=0}^3\sqrt{\lambda_{H,k}}\right)^2p(\eta)\mathrm{d}\eta.
\end{equation}

\section{QPSK Modulated CV-QKD with Multi-Stage CD-Kennedy Receiver and Post-Selection in Turbulent Channel}\label{CV-QKD_with_CD-Kennedy}
The CV-QKD protocols usually employ coherent states to distribute secret keys between two legitimate users Alice and Bob over a public channel in the presence of potential eavesdropper Eve. In this section, we consider a QPSK modulated CV-QKD protocol using multi-stage CD-Kennedy receiver based on post-selection strategy in turbulent channels.

In a QPSK modulated CV-QKD protocol in turbulent channels, Alice sends coherent signal $\ket{\beta_k}$ with equal priori probability to Bob through a turbulent channel. The legitimate channel between Alice and Bob is a turbulence deteriorated photon loss channel with channel transmission $\eta$ randomly varies according to a given lognormal distribution. Suppose the potential eavesdropper Eve can adopt the beamsplitter collective attack to obtain information about the key \cite{heid2006efficiency,sych2010coherent}. Then in a slow-varying turbulent channel, Eve can split $(1-\eta_0)$ quantity of the beam energy without being discovered. Then the quantum state arriving at Bob is $\ket{\beta_{B}}=\ket{\sqrt{\eta}\beta_k}$, where $\eta$ satisfies a lognormal distribution with average transmission $\eta_0$; and the quantum state arriving at Eve is $\ket{\beta_{E}}=\ket{\sqrt{1-\eta_0}\beta_k}$.

When a direct reconciliation is used, Bob detects the received state $\ket{\beta_{B}}$ and corrects his symbols according to Alice's revealed date. Then the SKR under collective attack is bounded by $S_{kr}\geq I_{AB}-\chi_E$, where $I_{AB}$ is the mutual information between Alice and Bob when a multi-stage CD-Kennedy receiver is employed; $\chi_E$ is the Holevo information of Eve's state $\ket{\beta_{E}}$.

\subsection{Mutual Information $I_{AB}$ Between Alice and Bob in Turbulent Channel}
\subsubsection{$I_{AB}$ with Multi-Stage CD-Kennedy Receiver}
To derive the mutual information $I_{AB}$ between Alice and Bob, we divide the channel into many effective information channels characterized by the parameters $\bm{n}$ and $\eta$ with
\begin{equation}
I_{AB}=\int_{\eta}\sum_{\bm{n}}p(\eta)p(\bm{n}|\eta,\bm{\gamma})I_{AB}(\bm{n},\eta)\mathrm{d}\eta,
\end{equation}
\noindent where $I_{AB}(\bm{n},\eta)$ is the effective information under the detected number of photons $\bm{n}\triangleq \{n_0,n_1,\cdots,n_{N-1}\}$ and the measured transmittance $\eta$ given displacement values $\bm{\gamma}$; $p(\bm{n}|\eta,\bm{\gamma})$ is the probability of detecting $\{n_0,n_1,\cdots,n_{N-1}\}$ photons at $N$ branches given transmittance $\eta$ and displacement values $\bm{\gamma}$, which can be obtained as
\begin{equation}\label{p_n_eta}
\begin{aligned}
p(\bm{n}|\eta,\bm{\gamma})&=\sum_{k=0}^3p_k p(\bm{n}|\beta_k,\eta,\bm{\gamma}),
\end{aligned}
\end{equation}
\noindent where $p(\bm{n}|\beta_k,\eta,\bm{\gamma})$ is given in \eqref{p_n_beta_k_eta}.

The effective information $I_{AB}(\bm{n},\eta)$ can be obtained as
\begin{equation}\label{I_AB_n_eta_beta_temp}
\begin{aligned}
&I_{AB}(\bm{n},\eta)\\
&\quad =H(A)-H(A|B)\\
&\quad =\sum_{k=0}^3p(\beta_k|\bm{n},\eta,\bm{\gamma})\log_2 p(\beta_k|\bm{n},\eta,\bm{\gamma})-\sum_{k=0}^3 p_k \log_2 p_k\\
&\quad =\sum_{k=0}^3 \frac{p_k p(\bm{n}|\beta_k,\eta,\bm{\gamma})}{p(\bm{n}|\eta,\bm{\gamma})}\log_2 \frac{p_k p(\bm{n}|\beta_k,\eta,\bm{\gamma})}{p(\bm{n}|\eta,\bm{\gamma})}\\
&\quad\quad \quad -\sum_{k=0}^3 p_k \log_2 p_k,
\end{aligned}
\end{equation}
\noindent where $H(x)$ is the entropy of $x$.

Substituting \eqref{p_n_beta_k_eta} and \eqref{p_n_eta} into \eqref{I_AB_n_eta_beta_temp}, we can obtain
\begin{equation}\label{I_AB_n_eta_beta}
\begin{aligned}
I_{AB}(\bm{n},\eta)&=\sum_{k=0}^3\frac{p_{k}\prod_{m=0}^{N-1} p(n_{m}|\beta_k,\eta,\gamma_m)}{\sum_{k=0}^3 p_{k} \prod_{m=0}^{N-1} p(n_{m}|\beta_k,\eta,\gamma_m)}\\
&\quad\quad \times\log_2 \frac{p_{k}\prod_{m=0}^{N-1} p(n_{m}|\beta_k,\eta,\gamma_m)}{\sum_{k=0}^3 p_{k} \prod_{m=0}^{N-1} p(n_{m}|\beta_k,\eta,\gamma_m)}\\
&\quad\quad\quad\quad -\sum_{k=0}^3 p_k \log_2 p_k,
\end{aligned}
\end{equation}
\noindent where $\gamma_0, \gamma_1, \cdots, \gamma_{N-1}$ are chosen according to the type of multi-stage CD-Kennedy receivers.

\subsubsection{$I_{AB}$ with Classical Coherent Receiver}
For comparison, here we give the mutual information between Alice and Bob when a classical coherent receiver is employed to receive the QPSK signals, i.e.,
\begin{equation}
I_{AB}^{HD}=\int_{\eta}\int_0^\infty\int_0^{2\pi} p(\eta) p(r,\theta|\eta)I_{AB}^{HD}(r,\theta,\eta)r \mathrm{d}\eta\mathrm{d}r\mathrm{d}\theta,
\end{equation}
\noindent where $I_{AB}^{HD}(r,\theta,\eta)$ is the effective information given output $(r,\theta)$ and transmittance $\eta$; $p(r,\theta|\eta)$ is the probability of output $(r,\theta)$ given transmittance $\eta$, which can be expressed as
\begin{equation}
\begin{aligned}
p(r,\theta|\eta)&=\sum_{k=0}^3 p_k p(r,\theta|\beta_k,\eta)\\
&=\sum_{k=0}^3 p_k \frac{1}{\pi} e^{-|re^{i\theta}-\sqrt{\eta}\beta_k|^2},
\end{aligned}
\end{equation}
\noindent where $p(r,\theta|\beta_k,\eta)$ is the PDF of detecting $(r,\theta)$ given transmitted signal $\beta_k$ and transmittance $\eta$, i.e.,
\begin{equation}
p(r,\theta|\beta_k,\eta)=\frac{1}{\pi} e^{-|re^{i\theta}-\sqrt{\eta}\beta_k|^2}.
\end{equation}

Similar to the multi-stage CD-Kennedy receiver, the effective information $I_{AB}^{HD}(r,\theta,\eta)$ for classical coherent receiver can be obtained as
\begin{equation}\label{I_AB_HD_r_theta_eta}
\begin{aligned}
&I_{AB}^{HD}(r,\theta,\eta)\\
&\quad =\sum_{k=0}^3p(\beta_k|r,\theta,\eta)\log_2 p(\beta_k|r,\theta,\eta)-\sum_{k=0}^3 p_k \log_2 p_k\\
&\quad =\sum_{k=0}^3\frac{p_{k}p(r,\theta|\beta_k,\eta)}{\sum_{k=0}^3 p_{k} p(r,\theta|\beta_k,\eta)}\log_2 \frac{p_{k}p(r,\theta|\beta_k,\eta)}{\sum_{k=0}^3 p_{k} p(r,\theta|\beta_k,\eta)}\\
&\quad\quad\quad\quad -\sum_{k=0}^3 p_k \log_2 p_k.
\end{aligned}
\end{equation}

\subsection{Holevo Information $\chi_E$}
The best eavesdropping strategy of Eve in a channel with negligible excess noise is the passive beamsplitter attack \cite{heid2006efficiency,sych2010coherent}. In a turbulent channel, Eve can split the beam near the transmitter and thus obtain $(1-\eta_0)$ quantity of the beam energy without being discovered. Then Eve has to discriminate the following QPSK modulated coherent states:
\begin{equation}
\hat{\rho}_k \triangleq \ket{\sqrt{1-\eta_0}\beta_k}\bra{\sqrt{1-\eta_0}\beta_k}, \quad k=0,1,2,3.
\end{equation}

Then in the view of Eve, the transmitted state is the following mixed state:
\begin{equation}
\hat{\rho}_E=\sum_{k=0}^3p_k\ket{\sqrt{1-\eta_0}\beta_k}\bra{\sqrt{1-\eta_0}\beta_k},
\end{equation}
\noindent where $p_k$ is the prior probabilities of transmitting symbol $k$.

According to the Holevo's theorem, the mutual information $I_{AE}$ between Alice and Eve is bounded by
\begin{equation}
I_{AE}\leq \chi_E\triangleq S(\hat{\rho}_E)-\sum_{k=0}^3 p_k S(\hat{\rho}_k),
\end{equation}
\noindent where $S(\hat{\rho})=-\text{Tr}(\hat{\rho})\log_2\hat{\rho}$ is the quantum entropy or von Neumann entropy of a quantum state $\hat{\rho}$.

For the pure state $\hat{\rho}_k$, we have $S(\hat{\rho}_k)=0$. Therefore, when $p_0=p_1=p_2=p_3=\frac{1}{4}$, we can obtain the Helevo information as (see Appendix \ref{Appendix I_AE})
\begin{equation}\label{I_AE}
\chi_E = -\sum_{k=0}^3 \lambda_k \log_2\lambda_k,
\end{equation}
\noindent where $\lambda_k$ with $k=0,1,2,3$ are given by
\begin{equation}\label{lambda_k}
\begin{cases}
\lambda_0=\frac{1-\eta_0}{2}e^{-N_s}(\cosh N_s+\cos N_s);\\
\lambda_1=\frac{1-\eta_0}{2}e^{-N_s}(\sinh N_s-\sin N_s);\\
\lambda_2=\frac{1-\eta_0}{2}e^{-N_s}(\cosh N_s-\cos N_s);\\
\lambda_3=\frac{1-\eta_0}{2}e^{-N_s}(\sinh N_s+\sin N_s).\\
\end{cases}
\end{equation}

\subsection{SKR of Multi-Stage CD-Kennedy Receiver with Post-Selection Strategy in Turbulent Channel}
Bob can use the post-selection strategy to achieve advantages over Eve. Specifically, Bob only save those bits with $I_{AB}(\bm{n},\eta)\geq \chi_E$ and discard the rest of bits. We define the post-selection area $\mathbf{A_{ps}}$ as
\begin{equation}
\mathbf{A_{ps}}=\{(\bm{n},\eta)|I_{AB}(\bm{n},\eta)\geq \chi_E\}.
\end{equation}

Then the secret key rate of the multi-stage CD-Kennedy receiver can be obtained as
\begin{equation}\label{S_kr_beta_1}
\begin{aligned}
S_{kr}&=\int_{\mathbf{A_{ps}}} p(\eta) p(\bm{n}|\eta) (I_{AB}(\bm{n},\eta)- \chi_E) \mathrm{d}\eta.
\end{aligned}
\end{equation}

\subsection{SKR of Classical Coherent Receiver with Post-Selection Strategy in Turbulent Channel}
Here we present the post-selection strategy for classical coherent receiver in turbulent channel. Similar to the multi-stage CD-Kennedy receivers, Bob can save those bits with $I_{AB}^{HD}(r,\theta,\eta)\geq \chi_E$, which corresponds to a post-selection area $\mathbf{A_{ps}^{HD}}$ defined as
\begin{equation}
\mathbf{A_{ps}^{HD}}=\{(r,\theta,\eta)|I_{AB}^{HD}(r,\theta,\eta)\geq \chi_E\}.
\end{equation}

Then the secret key rate of classical coherent receiver can be obtained as
\begin{equation}\label{S_kr_beta_heterodyne_1}
\begin{aligned}
S_{kr}^{HD}&=\int_{\mathbf{A_{ps}^{HD}}}  p(\eta)  p(r,\theta|\eta)(I_{AB}^{HD}(r,\theta,\eta)- \chi_E) r \mathrm{d}\eta\mathrm{d}r\mathrm{d}\theta.
\end{aligned}
\end{equation}

\section{Numerical Results}\label{Numerical_Results}
In this section we present some numerical results to verify both the error probability and SKR performance of proposed multi-stage CD-Kennedy receiver in turbulent channels. We define the signal intensity arriving at the receiver as $N_r\triangleq \eta_0 N_s=N N_0$. The turbulent strengths are represented by the turbulent variances and we use $\sigma_t^2=0.01$, $\sigma_t^2=0.1$, $\sigma_t^2=1$ to represent the weak turbulence, moderate turbulence, and strong turbulence, respectively. Unless otherwise specified, we set the signal photons $N_s=10$, the average transmittance $\eta_0=0.5$, the turbulent strength $\sigma_t^2=0.1$, and the number of stages $N=8$.

\begin{figure}
\centering
\includegraphics[width=0.7\textwidth]{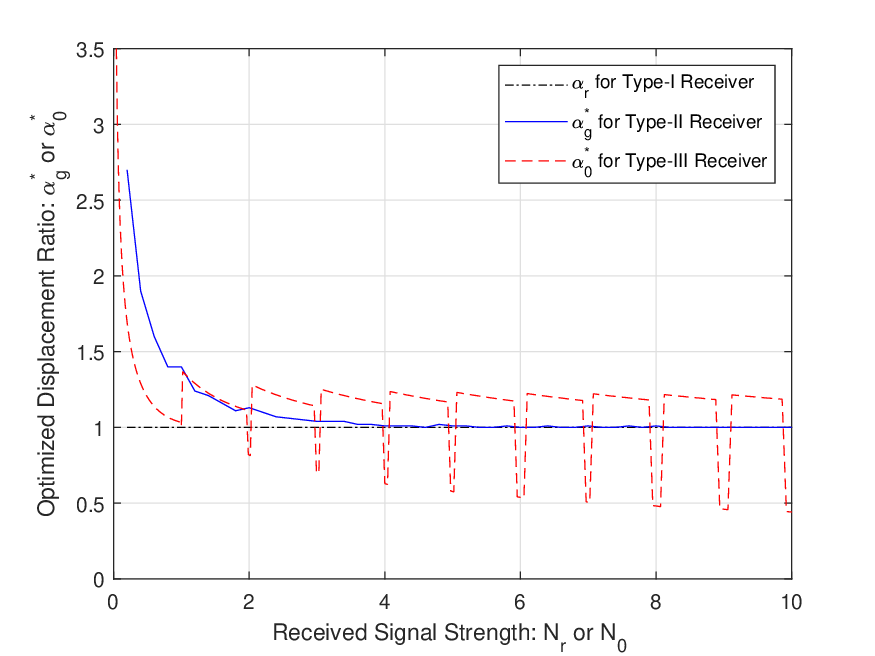}
\caption{Displacement ratios of different types of multi-stage CD-Kennedy receivers}
\label{displacement_ratios}
\end{figure}

We first present the optimized displacement ratios $\alpha_g^*$ and $\alpha_0^*$ for Type-II and Type-III receivers in Fig. \ref{displacement_ratios}, where the displacement ratio $\alpha_r=1$ for Type-I receiver is also plotted. From Fig. \ref{displacement_ratios} we can see that the optimized displacement ratio $\alpha_g^*$ decreases fast as the received signal strength $N_r$ increases when $N_r <1$ and will finally converge to 1 when $N_r$ is large. This indicates that the Type-II receiver will degrade to the Type-I receiver when the received signal strength is large. On the other hand, we can also see that the optimized displacement ratio $\alpha_0^*$ for Type-III receiver also decreases fast as the received signal strength $N_0<1$, but will not converge to 1 when the received signal strength is large. This indicates that the Type-III receiver will not degrade to the Type-I receiver when the received signal strength is large.

\begin{figure}
\begin{center}
\subfigure[]{\includegraphics[width=0.7\textwidth]{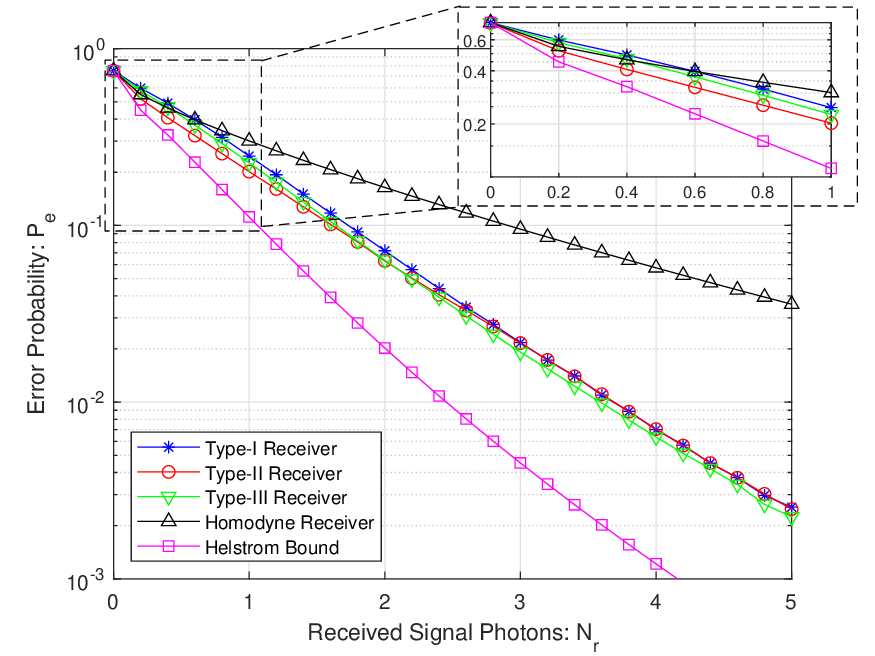}\label{Fig:Error_Prob_different_receivers}}
\subfigure[]{\includegraphics[width=0.7\textwidth]{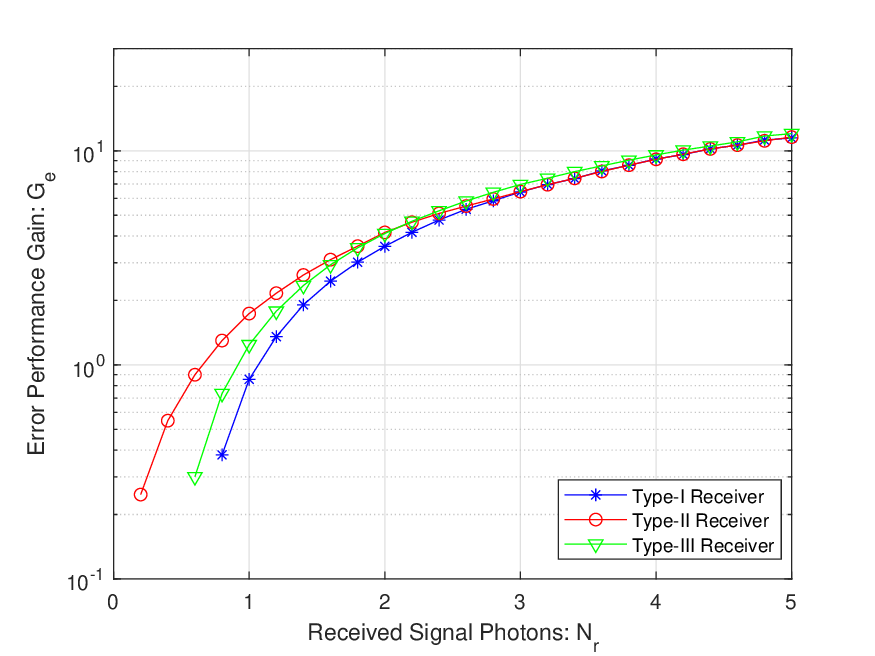}\label{Fig:Error_Prob_different_receivers_gain}}
\caption{Error probability performance comparison between different receivers: (a) error probability under different receivers; (b) error performance gains under different receivers}
\label{Fig:error_probability_comparison_different_receivers}
\end{center}
\end{figure}

Then we compare the error probability performance between different types of multi-stage CD-Kennedy receivers and homodyne receiver in Fig. \ref{Fig:Error_Prob_different_receivers}, where the Helstrom bound is also plotted. To better demonstrate the error probability performance advantage of different types of multi-stage CD-Kennedy receiver over the homodyne receiver, we can define the error performance gain as
\begin{equation}
G_e\triangleq 10\times \lg \frac{P_e^{HD}}{P_e},\\
\end{equation}
\noindent where $P_e$ and $P_e^{HD}$ are the error probabilities of multi-stage CD-Kennedy receiver and homodyne receiver, respectively. Then we also present the  error performance gain of different types of multi-stage CD-Kennedy receiver over the homodyne receiver in Fig. \ref{Fig:Error_Prob_different_receivers_gain}. From Fig. \ref{Fig:Error_Prob_different_receivers} we can observe that both the Type-II and Type-III receivers enjoys better error probability performance compared with the Type-I receiver. However, as we have analyzed above, the error of Type-II receiver degrades to that of the Type-I receiver, while the Type-III receiver enjoys better error probability performance than that of the Type-I receiver under all ranges of received signal strengths. Besides, we can also see that, almost all types of multi-stage CD-Kennedy receiver outperform the SQL of homodyne receiver, except for Type-I receiver with signal strength smaller than 0.6. Additionally, from Fig. \ref{Fig:Error_Prob_different_receivers_gain} we can also observe that the Type-II receiver enjoys the best error probability performance when the received signal strength is small, while the Type-III receiver enjoys the best error probability performance when the received signal strength is large.

\begin{figure}
\begin{center}
\subfigure[]{\includegraphics[width=0.7\textwidth]{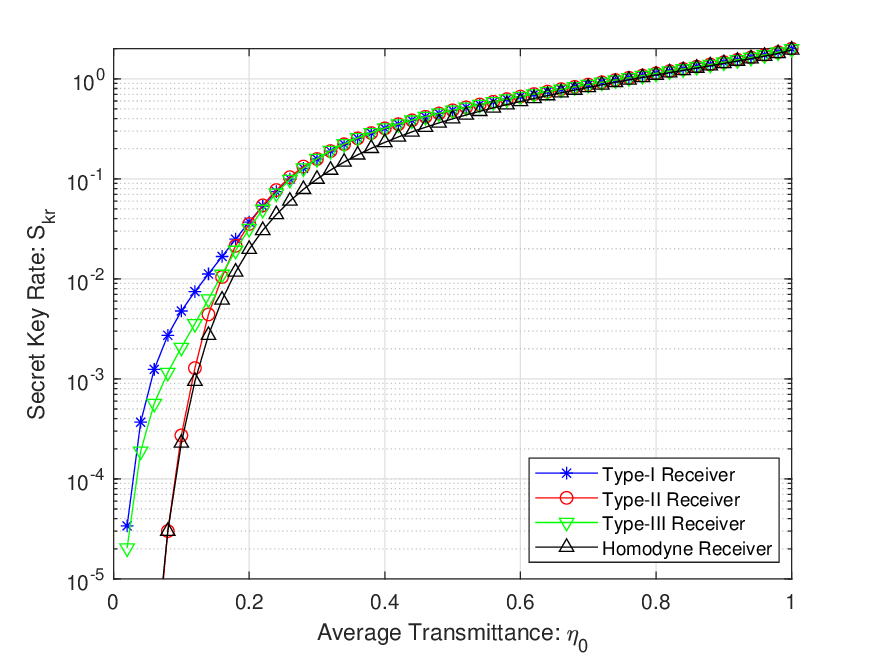}\label{Fig:SKR_different_receivers}}
\subfigure[]{\includegraphics[width=0.7\textwidth]{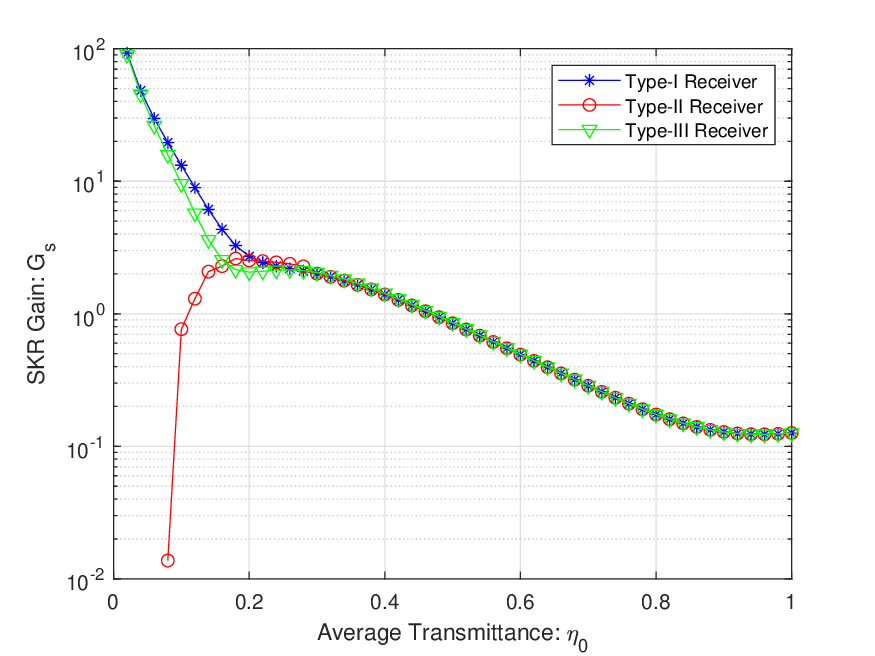}\label{Fig:SKR_different_receivers_gain}}
\caption{SKR performance comparison between different receivers: (a) SKR performance under different receivers; (b) SKR performance gain under different receivers}
\label{Fig:SKR_comparison_different_receivers}
\end{center}
\end{figure}

Then we present the SKR performance comparison between different receivers in Fig. \ref{Fig:SKR_different_receivers}. To better demonstrate the SKR performance advantage of different types of multi-stage CD-Kennedy receiver over the homodyne receiver, we can also define the SKR gain as
\begin{equation}
G_s\triangleq 10\times \lg \frac{S_{kr}}{S^{HD}_{kr}},\\
\end{equation}
\noindent where $S_{kr}$ and $S^{HD}_{kr}$ are the SKR of multi-stage CD-Kennedy receiver and homodyne receiver, respectively. Then we also present the SKR gain of different types of multi-stage CD-Kennedy receiver over the homodyne receiver in Fig. \ref{Fig:SKR_different_receivers_gain}. From Fig. \ref{Fig:SKR_different_receivers} we can observe that all types of multi-stage CD-Kennedy receiver can outperform the homodyne receiver in SKR performance. Besides, from Fig. \ref{Fig:SKR_different_receivers_gain} we can also see that,  when $\eta_0$ is small, the Type-I receiver enjoys the best SKR performance, then the Type-III receiver, and then the Type-II receiver; and the SKR of Type-II receiver will gradually approach that of the homodyne receiver. If we consider the performance results in both Figs. \ref{Fig:error_probability_comparison_different_receivers} and \ref{Fig:SKR_comparison_different_receivers}, we can conclude that, when the received signal strength is small, we should choose Type-II receiver for maximizing error probability performance and choose Type-I receiver for maximizing SKR performance; when the received signal strength is large, we should choose the Type-III receivers for maximizing both the error probability and SKR performance. Therefore, Type-III receiver becomes a good trade-off choice between error probability and SKR performance in all range of received signal strengths.

\begin{figure*}
\begin{center}
\subfigure[]{\includegraphics[width=0.48\textwidth]{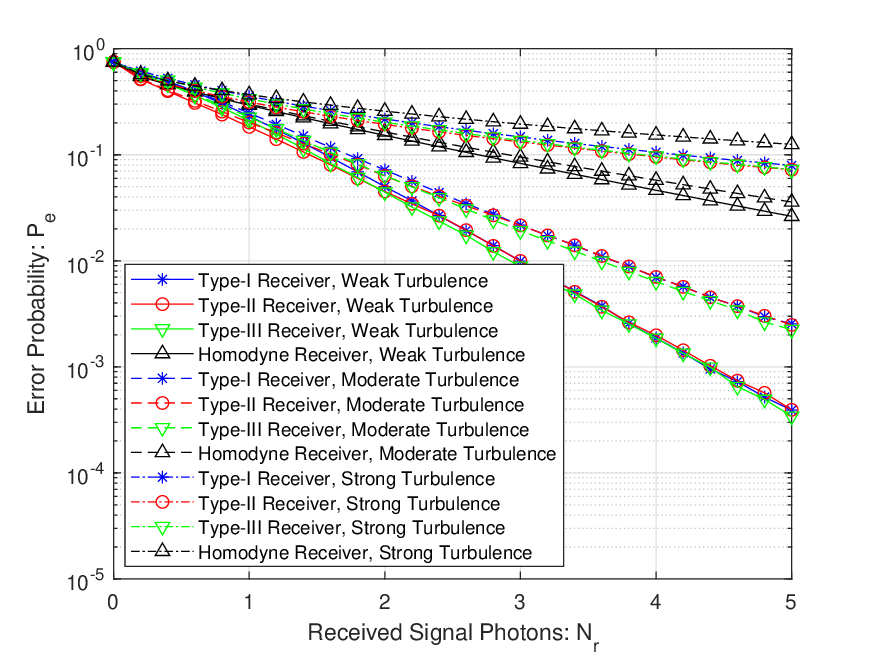}\label{Fig:Error_Prob_different_turbulence}}
\subfigure[]{\includegraphics[width=0.48\textwidth]{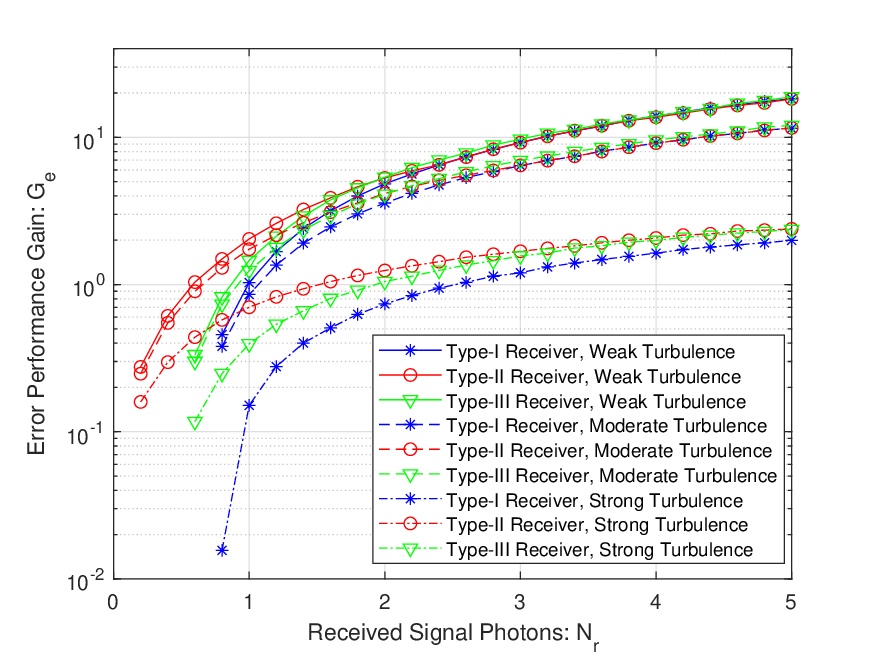}\label{Fig:Error_Prob_different_turbulence_gain}}
\subfigure[]{\includegraphics[width=0.48\textwidth]{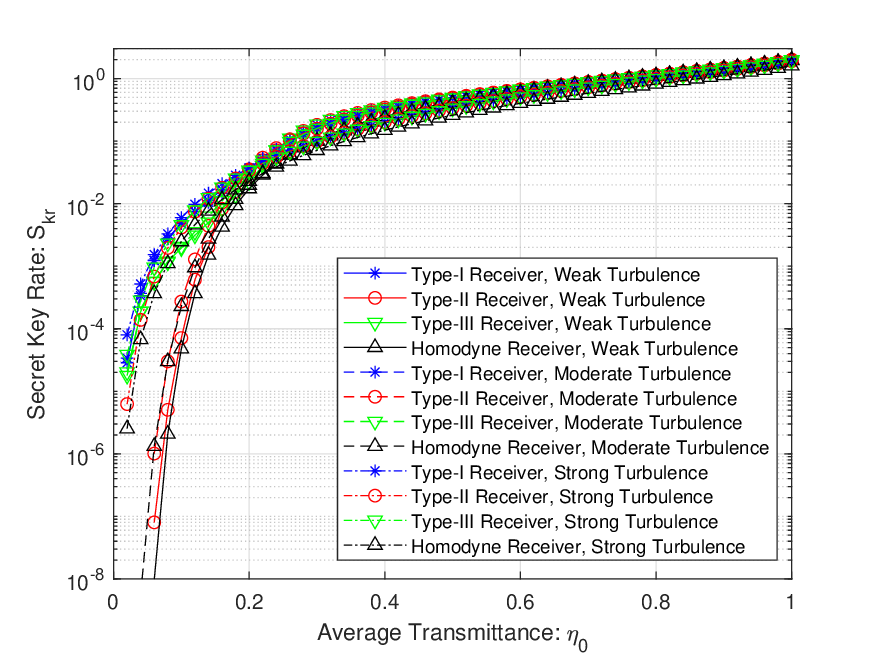}\label{Fig:SKR_different_turbulence}}
\subfigure[]{\includegraphics[width=0.48\textwidth]{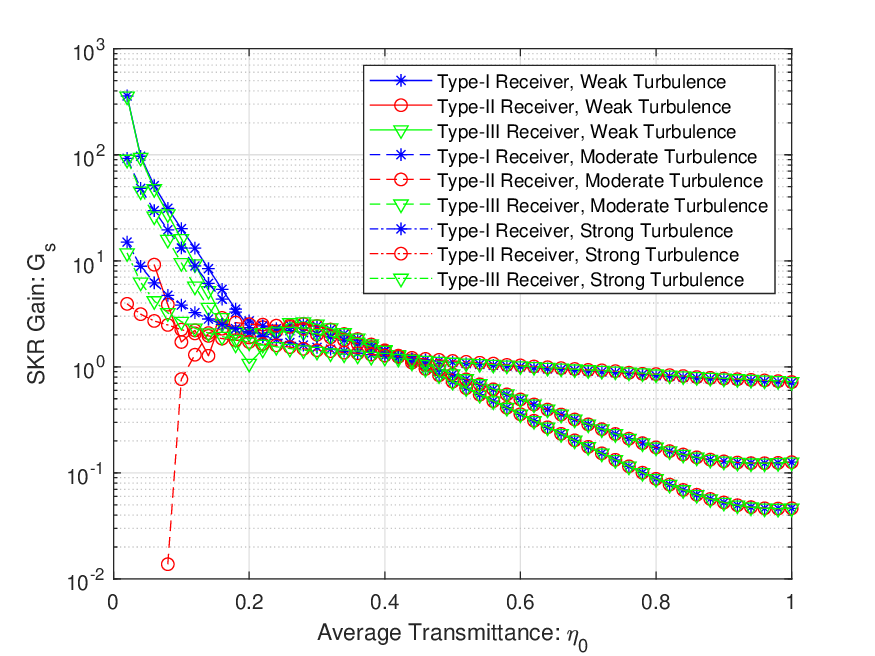}\label{Fig:SKR_different_turbulence_gain}}
\caption{Error probability and SKR performance comparison between different types of multi-stage CD-Kennedy receivers and homodyne receiver under different turbulent strengths: (a) error probability under different turbulent strengths; (a) error probability performance gain under different turbulent strengths; (c) SKR under different turbulent strengths; (d) SKR performance gain under different turbulent strengths}
\label{Fig:comparison_different_turbulence}
\end{center}
\end{figure*}

Then we compare the error probability and SKR performance under different turbulent strength $\sigma_t^2$ in Fig. \ref{Fig:comparison_different_turbulence}. From Figs. \ref{Fig:Error_Prob_different_turbulence} and \ref{Fig:Error_Prob_different_turbulence_gain} we can see that the error probability performance of the Type-II receiver is better than both Type-I and Type-III receivers when the received signal strength is small under different turbulent strengths. Besides, the advantage of Type-II receiver over Type-I and Type-III receivers will extend to large received signal strengths under strong turbulence. This indicates that the Type-II receiver can tolerate worse channel conditions compared with Type-I and Type-III receivers in terms of error probability performance. However, when the SKR performance is considered, the advantage of Type-II receiver vanishes under small received signal strengths, as shown in Figs. \ref{Fig:SKR_different_turbulence} and \ref{Fig:SKR_different_turbulence_gain}. Again, if both the error probability and SKR performance are considered, the Type-III receiver becomes a smart choice in all turbulent conditions.

\begin{figure*}
\begin{center}
\subfigure[]{\includegraphics[width=0.48\textwidth]{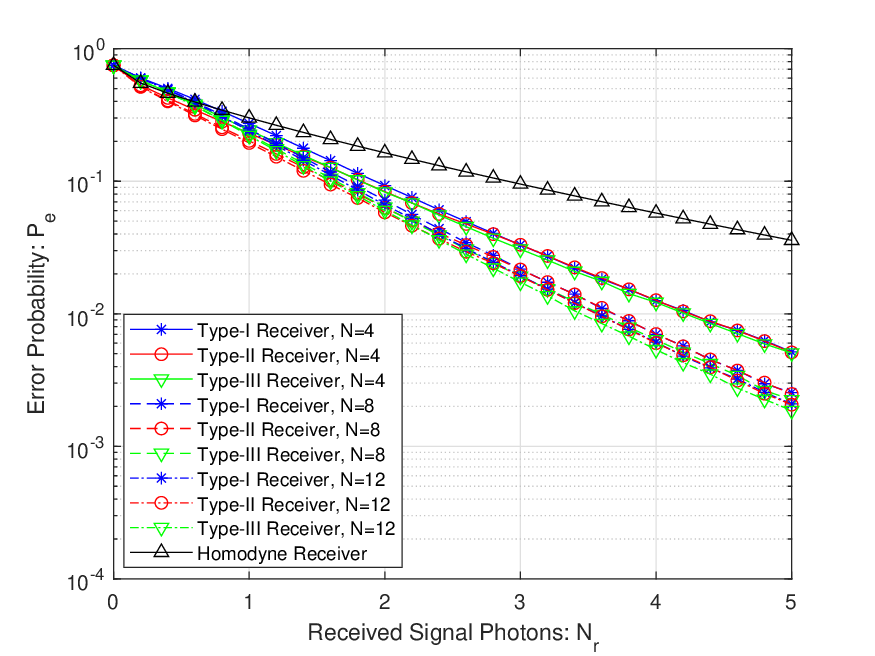}\label{Fig:Error_Prob_different_N}}
\subfigure[]{\includegraphics[width=0.48\textwidth]{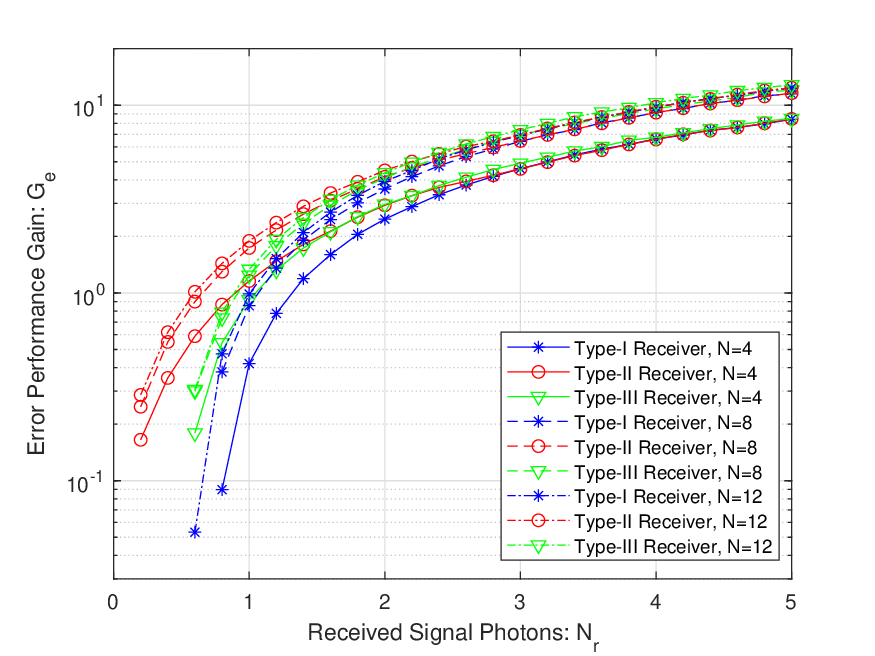}\label{Fig:Error_Prob_different_N_gain}}
\subfigure[]{\includegraphics[width=0.48\textwidth]{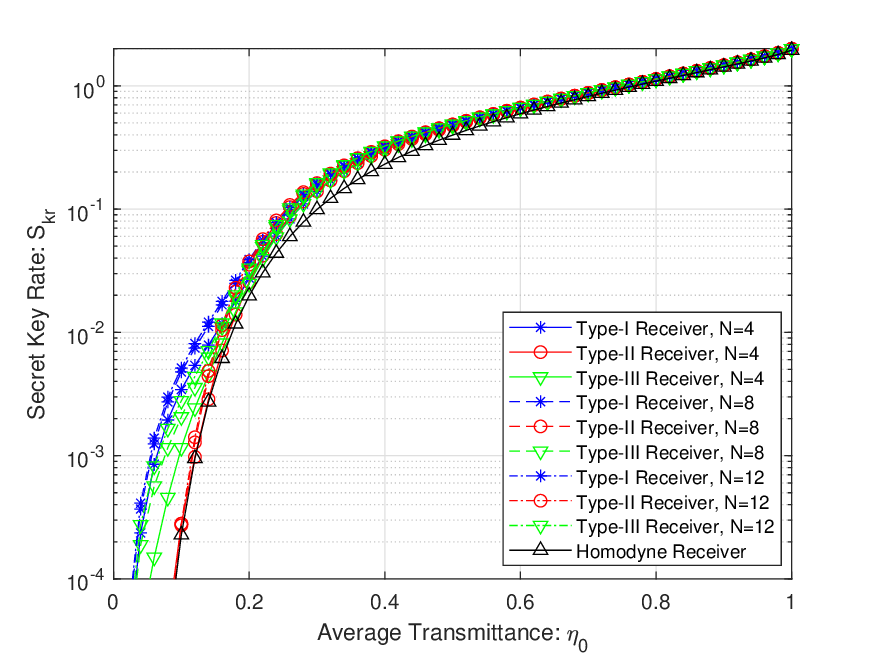}\label{Fig:SKR_different_N}}
\subfigure[]{\includegraphics[width=0.48\textwidth]{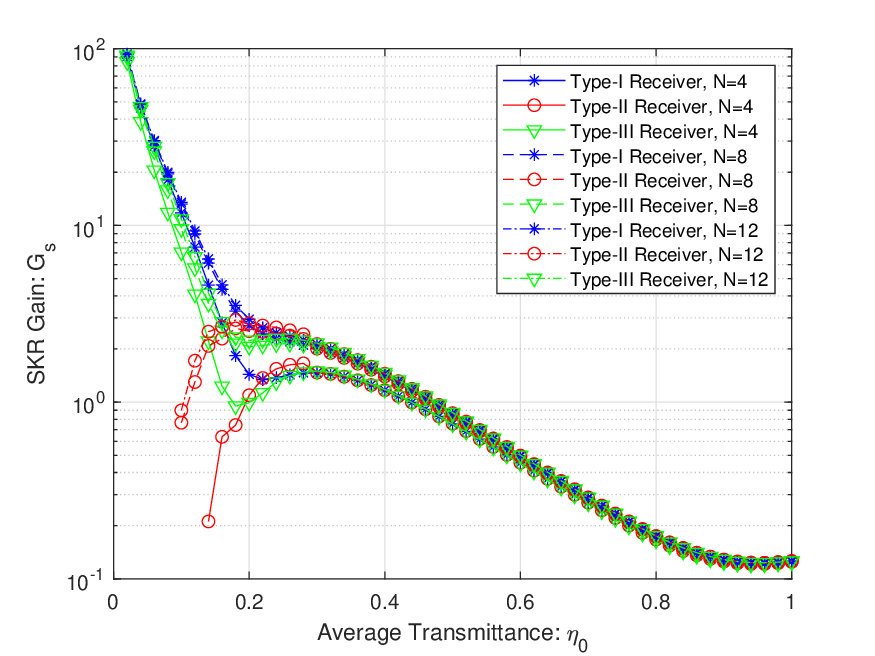}\label{Fig:SKR_different_N_gain}}
\caption{Error probability and SKR performance comparison between different types of multi-stage CD-Kennedy receivers and homodyne receiver under different number of stages: (a) error probability under different number of stages; (b) error probability performance gain under different number of stages; (c) SKR performance under different number of stages; (d) SKR performance gain under different number of stages}
\label{Fig:comparison_different_N}
\end{center}
\end{figure*}

At last, we present the error probability and SKR performance under different number of stages $N \in \{4, 8, 12\}$ in Fig. \ref{Fig:comparison_different_N}. From Fig. \ref{Fig:comparison_different_N} we can observe that both the error probability and SKR performance of all types of multi-stage CD-Kennedy receiver will improve as the number of stages $N$ increases. However, we can also see that though the error performance again of using multi-stage CD-Kennedy receiver over classical homodyne receiver will increase significantly as the received signal strength increases, the SKR gain of the multi-stage CD-Kennedy receiver only has incremental increase when the average transmittance $\eta_0>0.5$, as shown in Figs. \ref{Fig:SKR_different_N} and \ref{Fig:SKR_different_N_gain}. Besides, we can also observe that Type-III receiver enjoys a relative good performance of both error probability and SKR performance in all range of signal strengths. Additionally, if we consider both the error probability and SKR performance, we can conclude that the Type-III receiver can provide significant improvements in both error probability and SKR gains over the classical homodyne receiver when the received signal strength is small, i.e, in weak links.

\section{Conclusion}\label{Conclusion}
In this paper, we have thoroughly investigated the application of a multi-stage CD-Kennedy receiver to enhance the performance of a QPSK-modulated CV-QKD system operating over atmospheric turbulent channels. The primary limitation of classical coherent receivers, bound by the SQL, is particularly pronounced under the fluctuating conditions of satellite-to-ground links, necessitating the exploration of advanced detection schemes capable of surpassing this fundamental limit. To this end, we designed and analyzed three types of multi-stage CD-Kennedy receiver, i.e., the Type-I, Type-II, and Type-III receivers, with different strategy for applying displacement operators. We first derived the error probability of the multi-stage CD-Kennedy in the presence of channel transmittance fluctuations caused by turbulence. Subsequently, we integrated this receiver within a QPSK modulated CV-QKD protocol framework, incorporating a post-selection strategy, and derived the corresponding SKR. Numerical results conclusively demonstrate the significant superiority of the proposed multi-stage CD-Kennedy receiver over conventional homodyne receiver. The Type-II receiver exhibits remarkable resilience, achieving the lowest error probability and tolerating the most severe channel conditions among the three types. Meanwhile, the Type-III receiver emerges as a robust and versatile candidate, offering an excellent trade-off between error probability and SKR across the entire range of received signal strengths, with its advantages being especially pronounced in weak-link scenarios. This study establishes the multi-stage CD-Kennedy receiver as a potent technology for practical, high-performance CV-QKD in turbulent channels and shed a light on a more secure and efficient space-air-ground integrated network.

\bibliographystyle{IEEEtran}
\bibliography{ref}

\appendices

\section{Derivation of Error Probability $P_e(\eta)$}\label{Appendix P_e}
From \eqref{k_estimate} we can obtain the output symbol $\tilde{k}$ of the final branch as
\begin{equation}\label{k_estimate_final}
\begin{aligned}
\tilde{k}&=\mathop{\arg\max}_{k} \ \frac{p_{k}\prod_{i=0}^{N-1} p(n_{i}|\beta_k,\eta,\gamma_i)}{\sum_{k=0}^3 p_{k} \prod_{i=0}^{N-1} p(n_{i}|\beta_k,\eta,\gamma_i)}.
\end{aligned}
\end{equation}

Then the decision area $D_k$ for symbol $k$ on the final output photon numbers $n_{N-1}$ can be written as $D_k\triangleq \{n_{N-1}|\tilde{k}=k\}$ and the error probability given transmittance $\eta$ and $\bm{\gamma}$ can be expressed as
\begin{equation}\label{P_e_eta_D_k}
\begin{aligned}
P_e(\eta,\bm{\gamma})&=1-\sum_{k=0}^3\sum_{n_{N-1}\in D_k}p(n_{N-1}|\eta,\bm{\gamma}) \\
&\quad\quad\quad\quad\quad\quad \times p_r(n_{N-1}\in D_k |\beta_k,\eta,\bm{\gamma}),
\end{aligned}
\end{equation}
\noindent where $p(n_{N-1}|\eta,\bm{\gamma})$ is the probability of detecting $n_{N-1}$ photons under transmittance $\eta$ and displacement values $\bm{\gamma}$; $p_r(n_{N-1}\in D_k |\beta_k,\eta,\bm{\gamma})$ is the correct detecting probability when $n_{N-1}$ photons are detected given $\beta_k$ is transmitted under transmittance $\eta$ and displacement values $\bm{\gamma}$; and $p_r(n_{N-1}\in D_k |\beta_k,\eta,\bm{\gamma})$ can be expressed as
\begin{equation}\label{P_r_n_correct}
\begin{aligned}
p_r(n_{N-1}\in D_k |\beta_k,\eta,\bm{\gamma})&= \frac{p_k p(n_{N-1}|\beta_k,\eta,\bm{\gamma})}{p(n_{N-1}|\eta,\bm{\gamma})}.
\end{aligned}
\end{equation}

Substituting \eqref{P_r_n_correct} into \eqref{P_e_eta_D_k}, we can obtain
\begin{equation}\label{P_e_eta_D_k_2}
\begin{aligned}
P_e(\eta,\bm{\gamma})=1-\sum_{k=0}^3\sum_{n_{N-1}\in D_k} p_k p(n_{N-1}|\beta_k,\eta,\bm{\gamma}).
\end{aligned}
\end{equation}

Because the case of output $n_{N-1}$ photons consists of all possible output combination $\bm{n}\triangleq \{n_0,n_1,\cdots,n_{N-1}\}$ with the final output equals $n_{N-1}$, the decision area $D_k$ on $n_{N-1}$ is equivalent to the decision area $Q_k\triangleq \{\bm{n}|\tilde{k}=k\}$ on all possible output $\bm{n}$. Suppose $p(\bm{n}|\beta_k,\eta,\bm{\gamma})$ is the probability of obtaining $\bm{n}$ output photons when $\beta_k$ is transmitted under transmittance $\eta$ and displacement values $\bm{\gamma}$, we can future rewrite the error probability in \eqref{P_e_eta_D_k_2} as \eqref{P_e_eta_mid}.

\section{Proof of the Theorem 1}\label{multi-threshold_rank_detection}
As shown in Fig. \ref{displacement}, we first calculate the signals strength of the displaced state as $N_{0,0}\triangleq |\psi_0+\gamma_0|^2$, $N_{1,0}\triangleq \psi_1+\gamma_0$, $N_{2,0}\triangleq \psi_2+\gamma_0$, and $N_{3,0}\triangleq \psi_3+\gamma_0$ for state $\ket{\psi_0+\gamma_0}$, $\ket{\psi_1+\gamma_0}$, $\ket{\psi_2+\gamma_0}$, and $\ket{\psi_3+\gamma_0}$, respectively. By substituting $\beta_k$ and $\gamma_0$ into \eqref{lambda_m} we can further obtain
\begin{equation}\label{N_012}
\begin{cases}
N_{0,0}=(1-\alpha_0)^2N_0,\\
N_{1,0}=N_{3,0}=(1+\alpha_0^2)N_0,\\
N_{2,0}=(1+\alpha_0)^2N_0,\\
\end{cases}
\end{equation}
\noindent where $N_0\triangleq \frac{\eta N_s}{N}$ is the input signal strength at the first stage.

Suppose the decision areas of symbols $k=0$, $k=1$, $k=2$, and $k=3$ are $D_0$, $D_1$, $D_2$, and $D_3$, respectively. Because $\alpha_0>0$, we have $N_{0,0}< N_{1,0}=N_{3,0}< N_{2,0}$. Then we have the probabilities $p(n_0|\beta_1,\eta,\gamma_0)=p(n_0|\beta_3,\eta,\gamma_0)$, which indicates that symbol 1 and 3 share the same decision areas, i.e., $D_1=D_3$. Therefore, when the detected photon numbers $n_0$ fall into $D_1$, we should randomly choose the decided symbol as 1 or 3.

Now we consider the MAP detection between symbol 0 and 1, where decision rule is given by
\begin{equation}
\frac{p(n_0|\beta_0,\eta,\gamma_0)}{p(n_0|\beta_1,\eta,\gamma_0)} \mathop{\gtreqless}  \limits_{1}^{0} 1.
\end{equation}
\noindent We can define a function $g(n_0)\triangleq \frac{p(n_0|\beta_0,\eta,\gamma_0)}{p(n_0|\beta_1,\eta,\gamma_0)}$. Substituting  \eqref{Equa:photon_distribution_given_eta_i} and \eqref{N_012} into $g(n_0)$ we can obtain
\begin{equation}
g(n_0)=\left(\frac{1+\alpha_0^2-2\alpha_0}{1+\alpha_0^2}\right)^{n_0}e^{2\alpha_0 N_0}.
\end{equation}
\noindent Because $0\leq \frac{1+\alpha_0^2-2\alpha_0}{1+\alpha_0^2}< 1$, $g(n_0)$ decreases as $n_0$ increases. Then there exists a threshold $n_{th,0}$ such that $g(n_0)\geq 1$ always hold when $n_0\leq n_{th,0}$. Obviously, this equals a threshold detection $n_0 \mathop{\lesseqgtr} \limits_{1}^{0} n_{th,0}$, where $n_{th,0}=\big\lfloor \frac{2\alpha_0N_0}{\ln (1+\alpha_0^2)-\ln(1-\alpha_0)^2} \big\rfloor$ is obtained by letting $p(n_0|\beta_0,\eta,\gamma_0)=p(n_1|\beta_0,\eta,\gamma_0)$.

Similarly, we consider the MAP detection between symbol 1 and 2, which equals another threshold detection  $n_0 \mathop{\lesseqgtr} \limits_{2}^{1} n_{th,1}$ with $n_{th,1}=\big\lfloor \frac{2\alpha_0N_0}{\ln(1+\alpha_0)^2-\ln (1+\alpha_0^2)} \big\rfloor$.

Therefore, if we can prove $n_{th,0}<n_{th,1}$, then the MAP detection on symbols $\{0,1,2,3\}$ will reduce to the multiple-threshold rank detection given in \eqref{rank_detection}, which is sufficient to prove
\begin{equation}\label{inequal_1}
\frac{2\alpha_0N_0}{\ln(1+\alpha_0^2)-\ln(1-\alpha_0)^2}<\frac{2\alpha_0N_0}{\ln(1+\alpha_0)^2-\ln (1+\alpha_0^2)}.
\end{equation}
Because $\alpha_0>0$, we have $\ln(1+\alpha_0^2)-\ln(1-\alpha_0)^2>0$ and $\ln(1+\alpha_0)^2-\ln (1+\alpha_0^2)>0$. Then we can rewrite \eqref{inequal_1} as
\begin{equation}\label{inequal_2}
\ln [(1+\alpha_0)^2(1-\alpha_0)^2]^2 <\ln (1+\alpha_0^2)^2.
\end{equation}
\noindent This equivalence obviously holds since $1-\alpha_0^2<1+\alpha_0^2$, which concludes our proof.

\section{Derivation of Helevo Information $\chi_E$}\label{Appendix I_AE}
Before we derive the mutual information between Alice and Eve, we first introduce the following lemma.

\begin{lemma}
For arbitrary matrices $\bm{A}$ and $\bm{B}$, the matrices $\bm{AB}$ and $\bm{BA}$ share the same non-zero eigenvalues.
\end{lemma}
\begin{proof}
Suppose $\lambda_i$ is an non-zero eigenvalue of $\bm{AB}$ and $\bm{v}_i$ is the corresponding eigenvector, then we have
\begin{equation}\label{EID}
\bm{AB} \bm{v}_i =\lambda_i \bm{v}_i.
\end{equation}

Multiplexing $\bm{B}$ on both sides of \eqref{EID}, we can obtain
\begin{equation}\label{EID_2}
\bm{BA}(\bm{B}\bm{v}_i)=\lambda_i (\bm{B}\bm{v}_i).
\end{equation}
From \eqref{EID_2} we can see that $\lambda_i$ is also an eigenvalue of $\bm{BA}$ corresponding to an eigenvector $\bm{B}\bm{v}_i$.
\end{proof}

The quantum entropy of the mixed state $\hat{\rho}_E$ can be equivalently expressed as
\begin{equation}
S(\hat{\rho}_E)=-\sum_{n=0}^\infty \lambda_n \log_2\lambda_n,
\end{equation}
\noindent $\{\lambda_n\}_{n=0}^\infty$ is the eigenvalues of $\hat{\rho}_E$. It is challenging to perform the eigenvalue decomposition of $\hat{\rho}_E$ in the Hilbert space $\mathbb{H}$ with Fock basis because $\hat{\rho}_E$ is an infinite dimensional matrix in $\mathbb{H}$. However, since $\hat{\rho}_E$ is a linear combination of four pure states, we know that the number of non-zero eigenvalues of $\hat{\rho}_E$ cannot exceed four. Therefore, we can consider the subspace $\mathbb{V}$ spanned by the four coherent states $\{\ket{\sqrt{1-\eta_0}\beta_k}\}_{k=0}^3$ and perform the eigenvalue decomposition in this subspace $\mathbb{V}$. To achieve this, we define a operator $\hat{K}$ in subspace $\mathbb{V}$ with its matrix element given by
\begin{equation}\label{K_ij}
K_{i,j} \triangleq \sqrt{p_i}\langle \sqrt{1-\eta_0}\beta_{i} | \sqrt{1-\eta_0}\beta_j \rangle \sqrt{p_j},
\end{equation}
\noindent where $i,j \in \{0,1,2,3\}$.

In the other hand, we can rewrite $\hat{\rho}_E$ as
\begin{equation}
\hat{\rho}_E=\sum_{k=0}^3 \left(\sqrt{p_k}\ket{\sqrt{1-\eta_0}\beta_k}\right)\left(\sqrt{p_k}\bra{\sqrt{1-\eta_0}\beta_k}\right).
\end{equation}

Then we can define two operators $\hat{A}$ and $\hat{B}$ as follows:
\begin{equation}
\begin{cases}
\hat{A}\triangleq \sum_{k=0}^3 \left(\sqrt{p_k}\ket{\sqrt{1-\eta_0}\beta_k}\right) \bra{v_k}\\
\hat{B}\triangleq \sum_{k=0}^3 \ket{v_k}\left(\sqrt{p_k}\bra{\sqrt{1-\eta_0}\beta_k}\right),
\end{cases}
\end{equation}
\noindent where $\{\ket{v_k}\}_{k=0}^3$ is a set of orthogonal basis of subspace $\mathbb{V}$. Then we can obtain $\hat{A}\hat{B}$ as
\begin{equation}
\begin{aligned}
\hat{A}\hat{B}&=\!\sum_{i=0}^3 \sum_{j=0}^3 \left(\!\sqrt{p_i}\ket{\sqrt{1-\eta_0}\beta_i}\!\right)\langle v_i|v_j\rangle \left(\!\sqrt{p_j}\bra{\sqrt{1-\eta_0}\beta_j}\!\right)\\
&=\sum_{k=0}^3 \left(\sqrt{p_k}\ket{\sqrt{1-\eta_0}\beta_k}\right)\left(\sqrt{p_k}\bra{\sqrt{1-\eta_0}\beta_k}\right)\\
&=\hat{\rho}_E.
\end{aligned}
\end{equation}

Similarly, we can obtain $\hat{B}\hat{A}$ as
\begin{equation}
\begin{aligned}
\hat{B}\hat{A}&=\sum_{i=0}^3\sum_{j=0}^3 \ket{v_i} \left(\sqrt{p_i}\langle \sqrt{1-\eta_0}\beta_{i} | \sqrt{1-\eta_0}\beta_j \rangle \sqrt{p_j}\right)\bra{v_j}\\
&=\sum_{i=0}^3\sum_{j=0}^3 \ket{v_i}K_{i,j}\bra{v_j}\\
&=\hat{K}.
\end{aligned}
\end{equation}

According to Lemma 1, $\hat{\rho}_E$ and $\hat{K}$ share the same non-zero eigenvalues. Therefore, we can express the quantum entropy $S(\hat{\rho}_E)$ as
\begin{equation}
S(\hat{\rho}_E)=-\sum_{k=0}^3 \lambda_k \log_2\lambda_k,
\end{equation}
\noindent where $\lambda_k$ is the eigenvalues of $\hat{K}$.

Substituting $\beta_k=\beta e^{i\frac{\pi}{4}(2k+1)}$ into \eqref{K_ij}, we can obtain $K_{ij}$ as
\begin{equation}
K_{i,j}=\sqrt{p_i p_j}(1-\eta_0)e^{-N_s(1- e^{i\frac{\pi}{2}(j-i)})}
\end{equation}

Specifically, when $p_0=p_1=p_2=p_3=\frac{1}{4}$, we have
\begin{equation}
\hat{K}=\frac{(1-\eta_0)e^{-N_s}}{4} \left[
\begin{matrix}
e^{N_s}& e^{iN_s} & e^{-N_s} &e^{-iN_s}\\
e^{-iN_s} & e^{N_s} & e^{iN_s} &e^{-N_s} \\
e^{-N_s} & e^{-iN_s} & e^{N_s} &e^{iN_s}\\
e^{-iN_s} & e^{-N_s} & e^{-iN_s} &e^{N_s}\\
\end{matrix}
\right].
\end{equation}

Since $\hat{K}$ is a circulant matrix, the eigenvalue decomposition of $\hat{K}$ can be performed using discrete Fourier transformation and the eigenvalues are given by
\begin{equation}
\begin{aligned}
\lambda_k&=\sum_{j=0}^3K_{0,j}(e^{i\frac{\pi}{2}})^{kj}.
\end{aligned}
\end{equation}

By substituting $K_{0,j}$ into above equation, we can further obtain all eigenvalues in \eqref{lambda_k}.

\balance

\end{document}